\documentclass[aps,prx,superscriptaddress,notitlepage,nopacs,citeautoscript,longbibliography,floatfix,letter,aps,twocolumn,10pt]{revtex4-2}
\usepackage{epsfig,amsmath,amssymb,color,comment,physics, dsfont, bbold}
\usepackage[makeroom]{cancel}
\usepackage[caption=false]{subfig}
\usepackage{mathrsfs}
\usepackage[countmax]{subfloat}
\usepackage[normalem]{ulem}
\usepackage[english]{babel}
\usepackage{dsfont}
\usepackage{float}
\usepackage{color}
\usepackage[dvipsnames]{xcolor}
\usepackage[bookmarks=true,colorlinks,linkcolor=blue,urlcolor=NavyBlue,citecolor=RoyalBlue]{hyperref}
\usepackage{orcidlink}

\begin{document}

\title{Tunable quantum Mpemba effect in long-range interacting systems
}

\author{Andrew Hallam \orcidlink{0000-0003-2288-7661}}
\thanks{These authors contributed equally.}
\affiliation{School of Physics and Astronomy, University of Leeds, Leeds LS2 9JT, United Kingdom}

\author{Matthew Yusuf \orcidlink{0009-0005-7080-8499}}
\thanks{These authors contributed equally.} 
\affiliation{School of Physics and Astronomy, University of Leeds, Leeds LS2 9JT, United Kingdom}

\author{Aashish A. Clerk}
\affiliation{Pritzker School of Molecular Engineering, University of Chicago, Chicago, Illinois 60637, USA}

\author{Ivar Martin}
\affiliation{Materials Science Division, Argonne National Laboratory, Lemont, Illinois 60439, USA}
\affiliation{Department of Physics, University of Chicago, Chicago, Illinois 60637, USA}

\author{Zlatko Papi\'c \orcidlink{0000-0002-8451-2235}}
\affiliation{School of Physics and Astronomy, University of Leeds, Leeds LS2 9JT, United Kingdom}

\date{November 10, 2026}
\begin{abstract}

Symmetry plays a fundamental role in many-body systems, both in and out of equilibrium. The quantum Mpemba effect (QME)---a phenomenon where systems initially farther from equilibrium can thermalize faster---can be understood in terms of how rapidly a symmetry, broken by initial conditions, is dynamically restored. In this work, we study the QME in a one-dimensional spin-1/2 XYZ model with power-law decaying interactions in the presence of a magnetic field. In the prethermal regime generated by large field strengths, the system develops a continuous U(1) symmetry, enabling the QME to emerge. However, due to the Hohenberg-Mermin-Wagner theorem, the QME can only arise when interactions are sufficiently short-ranged. This leads to an interplay between the external field, interaction range, and dynamical symmetry restoration. We systematically explore this interplay and analyze the dependence of the QME on the effective temperature set by the initial state. Our results demonstrate the tunability of the QME via long-range interactions, which can be probed in experimental platforms of trapped ions, polar molecules, and NV centers.

\end{abstract}
\maketitle

\section{Introduction}\label{sec:introduction}

The notion of spontaneous symmetry breaking (SSB) is central to equilibrium many-body physics. Within the Landau paradigm, phases of matter and transitions between them can be understood by their relationship to the symmetries of the underlying Hamiltonian.  Recently, attention has turned to understanding symmetry breaking in nonequilibrium quantum systems. According to the eigenstate thermalization hypothesis (ETH)~\cite{Deutsch1991,Srednecki1994}, for most Hamiltonians and most initial states, the system is expected to relax to thermal equilibrium~\cite{RigolNature,dAlessio2016,Ueda2020}. Without SSB, the resulting thermal state should be invariant under the symmetries of the system. Therefore, even if the initial state breaks these symmetries, one expects the symmetry to be restored during the time evolution. 

Somewhat surprisingly, in many cases it has been found that the \emph{more} symmetry-broken an initial state is, the \emph{more rapid} the symmetry restoration. By analogy with the similar phenomenon in water~\cite{Mpemba1969}, this behavior has been dubbed the quantum Mpemba effect (QME)~\cite{ares2025}. The QME has been predicted in a wide range of systems, including quantum spin chains \cite{Ares2023,Murciano2024,Liu2024_2}, lattice models of fermions and bosons~\cite{Yamashika2024,yamashika2024quenchingsuperfluidfreebosons}, quantum circuits \cite{ares2025entanglementasymmetrydynamicsrandom,yu2025,Liu2024_3,turkeshi2024,klobas2024, Klobas2024Rule54,Foligno2025}, quantum dots \cite{graf2025, Zatsaryna2025}, and open quantum systems \cite{Carollo2021, Zhang2025, Kochsiek2022, bao2022, Ivander2023, Zhou2023, Chatterjee2023, Wang2024, Liu2024_4, Longhi2024, wang2024mpembameetsquantumchaos, furtado2024, qian2024, Bettmann2025, Dong2025, Nava2024, medina2024, kheirandish2024,  strachan2024, wang2024goingquantummarkovianityreality, mondal2025, ma2025, wei2025, ali2025, Li2025, chatterjee2025}. Recently, the QME has also been realized in quantum simulation platforms~\cite{Joshi2024,xu2025}. 
Most of these works have focused on the restoration of continuous symmetries, in particular, U(1) symmetry associated with particle-number conservation, although systems with non-Abelian SU(2) symmetry have also been shown to exhibit the QME \cite{Rylands2024, Liu2024_3, Yu2025_2, fujimura2025}. In integrable models, the QME arises due to states that are simultaneously more out-of-equilibrium and have larger overlaps with the fastest quasiparticle modes~\cite{Rylands2024_2}. Similar mechanisms have been used to explain the effect in systems coupled to external reservoirs~\cite{Ares2025FreeFermionsMixed,Caceffo2024}. Recently, Ref.~\cite{bhore2025} argued that the QME can be generalized to chaotic systems without spatial symmetries or even energy conservation, where there is no quasiparticle description at high energies. This suggests the ubiquity of the QME in quantum many-body systems.

Despite being a nonequilibrium phenomenon, the QME depends upon the eventual restoration of symmetry, and is therefore intimately connected to the system's equilibrium properties. This motivates the question of how the QME and SSB interplay in the context of the Hohenberg-Mermin-Wagner (HMW) theorem~\cite{Mermin1966,Hohenberg1967}, a celebrated no-go theorem for SSB in spatial dimensions $d\leq 2$. One may expect the QME to occur if a phase does not exhibit SSB; conversely, the QME is not expected to persist in a phase with SSB. The existence of SSB can be tuned in systems with long-range (algebraically decaying) interactions \cite{Gong2016Kaleidoscope, Gong2016TopologicalPhases, Maghrebi2017, Defenu2023, Chen_2023, Feng_2023}, which can evade the HMW theorem even in $d=1$.  Power-law decaying interactions arise naturally in trapped ions, polar molecules, and NV centers in diamond \cite{Gross2017,Schafer2020,Choi2020,Zhou2020} (see also Ref.~\cite{Defenu2023} for a recent overview). In particular, trapped-ion experiments allow the effective decay exponent to be tuned from nearly all-to-all to dipolar regimes by adjusting laser detunings and phonon-mode couplings.

Long-range interactions are a well-known source of unusual nonequilibrium properties, such as suppressed growth of entanglement entropy and dynamical phases of matter~\cite{Lerose2020,Machado2020}. This raises a natural question: how does the QME depend on the range of interactions? A recent work~\cite{yamashika2025} investigated signatures of the QME in long-range spin systems using spin-wave analysis and showed that the QME can persist over a range of interaction decay exponents. However, the role of SSB and its implications for dynamical symmetry restoration were not addressed. In particular, it remains unclear whether sufficiently long-range interactions can qualitatively suppress the QME by stabilizing symmetry-broken phases, and how such behavior is modified in systems with only emergent rather than exact continuous symmetries.

Here we study the QME in a one-dimensional (1D) long-range interacting spin-$1/2$ XYZ chain subject to a longitudinal magnetic field. In the large-field prethermal regime, an emergent U(1) symmetry enables dynamical symmetry restoration, whereas sufficiently long-range interactions promote its SSB. Because our central interest lies in dynamical symmetry restoration, we consider models that do not possess an explicit U(1) symmetry, yet develop an emergent one in the rotating frame at large fields. This approach allows us to capture both symmetry-broken and symmetry-restoring dynamics within a unified framework. By varying the field strength, interaction range, and effective temperature of the initial state, we map out a ``QME dynamical phase diagram'' and identify the conditions under which the effect halts. These results connect the QME to the HMW theorem and establish tunable routes for its observation in quantum-simulation platforms.

The remainder of this paper is organized as follows. In Sec.~\ref{sec:longrange}, we introduce the XYZ model and briefly review the physics of its prethermal regime, with an emergent U(1) symmetry and a continuous time crystal phase. This regime will be key to stabilizing the QME.  In Sec.~\ref{sec:longrange} we also introduce the notion of entanglement asymmetry, $\Delta S_A$, which will be used as the main diagnostic of the QME. Our main results are presented in Sec.~\ref{sec:resXYZ} and~\ref{sec:resXXZ}, where we explore two directions in the dynamical phase diagram: varying the strength of the magnetic field and the range of interactions. In Sec.~\ref{sec:resXYZ}, we study the effect of varying the magnetic field strength, while in Sec.~\ref{sec:resXXZ} we focus on the large-field regime, studying the effect of U(1) SSB on the QME by varying the interaction range and the energy density of the initial state. The results of Sec.~\ref{sec:resXXZ} directly apply to static systems described by isotropic XXZ Hamiltonians. We summarize our findings in the QME dynamical phase diagram presented in Sec.~\ref{sec:conc}, while Supplemental Material (SM)~\cite{SM} provides further details of trace distance as an alternative measure of the QME, additional numerical results and finite-size scaling analysis, and the study of the SU(2)-symmetric point in the XYZ model.

\section{Long-range XYZ model}\label{sec:longrange}

\begin{figure}[t]
    \centering
    \includegraphics[width=\linewidth]{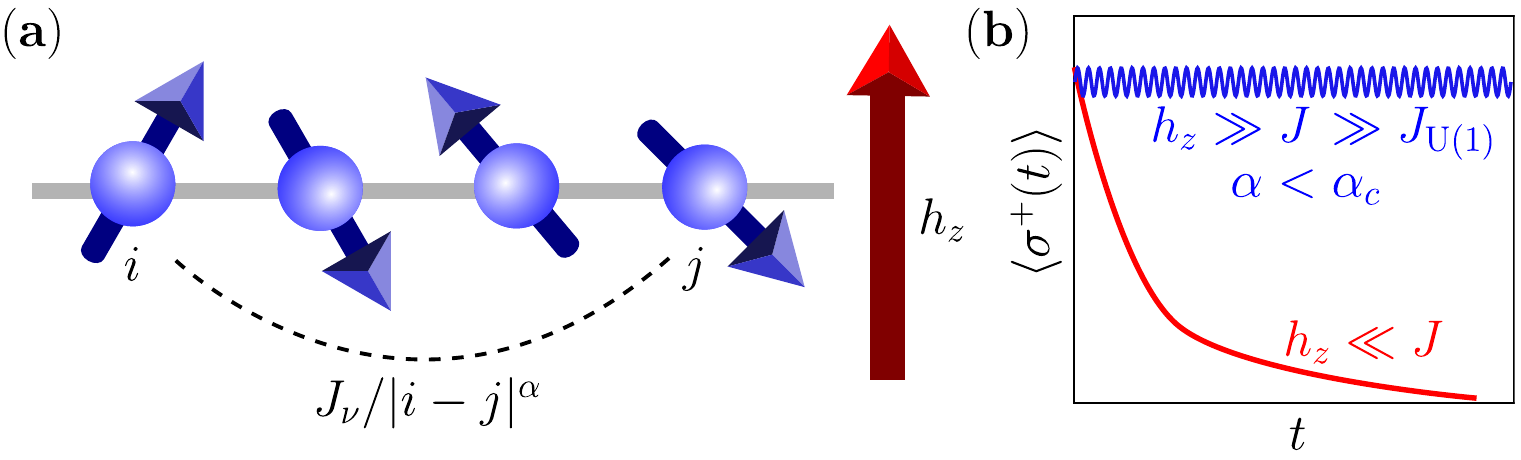}
    \caption{(a) Schematic of a one-dimensional long-range XYZ spin chain in an external magnetic field $h_z$. The spins $i$ and $j$ interact via power-law decaying interactions, $(J_\nu/|i-j|^\alpha) \hat \sigma^\nu_i\hat \sigma^\nu_j$, where $J_\nu$, $\nu=x,y,z$ are the interaction couplings along the three directions and $\hat \sigma_i^\nu$ are Pauli matrices, as in Eq.~(\ref{eq:Ham}). (b) At weak fields, $h_z\ll J_\nu$, there is no U(1) symmetry for generic values of $J_\nu$ and the corresponding order parameter, $\sigma^+ \equiv (\sigma^x + i \sigma^y)/2$, rapidly decays to zero as a function of time $t$. By contrast, in the prethermal regime when $h_z \gg J \gg J_\mathrm{U(1)} \equiv |J_x-J_y|$, where $J$ is the largest of the couplings $J_\nu$, the system develops a U(1) symmetry in the rotating frame, accompanied by the persistent nonzero value of $\langle \sigma^+(t)\rangle$---the hallmark of a continuous time crystal~\cite{Else2017}. This only occurs when $\alpha$ is smaller than a critical value $\alpha_c$ (see text for details).
    The goal of this paper is to quantitatively explore the competition between spontaneous symmetry breaking and dynamical symmetry restoration.}
    \label{fig:XYZmodel}
\end{figure}

Inspired by experimental platforms such as trapped ions and polar molecules~\cite{Defenu2023}, we consider a one-dimensional spin-$1/2$ long-range XYZ spin model in a magnetic field pointing along the $z$ direction, illustrated in Fig.~\ref{fig:XYZmodel}(a). The model is defined by the Hamiltonian
\begin{eqnarray}
    \label{eq:Ham}
\hat H  = \frac{1}{\mathcal{N}} \sum_{i<j} \sum_{\nu\in\{x,y,z\}} \frac{J_\nu}{|i-j|^\alpha}  \,\hat \sigma^\nu_i\hat \sigma^\nu_j 
+ h_z\sum_i \hat \sigma^z_i,
\end{eqnarray}
where $J_{\nu=x,y,z}$ are the spin-spin couplings, $\hat \sigma_i^\nu$ are the standard Pauli matrices on site $i$, and $h_z$ is the strength of the external magnetic field. The power-law exponent $\alpha$ sets the range of interactions. We assume open boundary conditions for the chain and normalize the spin-spin couplings using the Kac norm~\cite{Kac1963}, 
\begin{equation}
 \mathcal{N} = \frac{1}{N-1}\sum_{i\neq j} \frac{1}{|i-j|^\alpha},   
\end{equation}
where $N$ is the total number of spins. This normalization ensures that the energy density is intensive in the thermodynamic limit and allows for a consistent comparison of results at different $\alpha$. 

In the limit $J_x=J_y$, the Hamiltonian reduces to the long-range, U(1)-symmetric XXZ model. When $\alpha \rightarrow \infty$, interactions are restricted to nearest neighbors and the model becomes the quantum integrable XXZ spin chain~\cite{sutherland2004beautiful}. In the opposite limit $\alpha=0$, the Hamiltonian is a fully connected Lipkin-Meshkov-Glick (LMG) model~\cite{LMG1,LMG2}, which can be solved by mapping to a free collective spin~\cite{Sciolla2010,Sciolla2013,LerosePappalardi}. However, for any finite $\alpha$, the model~(\ref{eq:Ham}) is nonintegrable. 
Previous studies of ground-state properties at $h_z=0$ and finite $\alpha$~\cite{Maghrebi2017,Ren2020} showed that the XXZ model with $J_z \lessapprox
1$ transitions from a gapless, short-range U(1)-preserving XY phase (large $\alpha$), to a gapless, U(1)-breaking superfluid phase (small $\alpha$), undergoing a phase transition at a critical $\alpha_c\approx 3$. In the remainder of this section, we focus on the dynamics of the model~(\ref{eq:Ham}) in the regime of large fields, where it develops an emergent U(1) symmetry and time crystal phase. We also introduce the entanglement asymmetry, which will be used as the main diagnostic of the QME in subsequent sections.

\subsection{Prethermal U(1) symmetry and continuous time crystal}\label{sec:CTC}

While in equilibrium the model~(\ref{eq:Ham}) only exhibits U(1) symmetry when $J_x=J_y$, away from equilibrium it is possible to effectively realize U(1) symmetry for generic couplings $J_\nu$ if the field is large, i.e., $h_z\gg J$, where $J$ is the largest of the couplings $J_\nu$. In the rotating frame aligned with the $z$ axis,  there is an emergent U(1) symmetry, which can leave dynamical signatures when the system is quenched far from equilibrium, e.g., in the form of a continuous time crystal (CTC)~\cite{Else2017,Bull2022,Machado2020}. We now briefly review this scenario, summarized in Fig.~\ref{fig:XYZmodel}(b), as the same regime can be used to create the QME for arbitrary couplings $J_\nu$ in Eq.~(\ref{eq:Ham}), by varying $\alpha$ (and $h_z$ remaining sufficiently large). 

According to the static prethermal theorem~\cite{Abanin_2017}, a Hamiltonian of the form in Eq.~(\ref{eq:Ham}), i.e., $\hat H=\hat H_0 + h_z \hat Z$, where $\hat Z=\sum_i \hat \sigma_i^z$ has an integer spectrum, can be transformed (through a series of unitary rotations in the large-$h_z$ limit) into $\hat D + \hat V + h_z \hat Z$, where $\hat D$ commutes with $\hat Z$ and $\hat V$ is an exponentially small correction in $h_z$. For our model in Eq.~(\ref{eq:Ham}), this yields~\cite{Bull2022} 
\begin{equation}
    \hat D=\frac{1}{\mathcal{N}}\sum_{i<j}  \frac{J_x + J_y}{2} \left(\frac{\sigma_i^x \sigma_j^x}{\vert i-j \vert^{\alpha}} + \frac{\sigma_i^y \sigma_j^y}{\vert i-j \vert^{\alpha}}\right) + J_z \frac{\sigma_i^z \sigma_j^z}{\vert i-j \vert^{\alpha}}.  \label{eq:D}
\end{equation}
For timescales that are exponentially long in $h_z/J$ the dynamics of the model~(\ref{eq:Ham}) are governed by an effective prethermal Hamiltonian 
\begin{equation}\label{eq:XXZ}
\hat H_\mathrm{XXZ}^\mathrm{eff} \approx \hat D + h_z \hat Z, \quad \text{valid for} \quad  t \ll t_* \sim e^{O(h_z/J)}.    
\end{equation}
This effective Hamiltonian has a U(1) symmetry generated by $\hat Z$. Intuitively, Eq.~(\ref{eq:XXZ}) expresses the fact that the large $z$-magnetic field tends to ``homogenize'' interactions in the $x$ and $y$ directions in the rotating frame. 
Note that there are correction terms to Eq.~(\ref{eq:XXZ}) which depend on the explicit U(1)-breaking in the couplings, $J_\mathrm{U(1)}\equiv |J_x-J_y|$~\cite{Bull2022}. Thus the emergent U(1) symmetry in the prethermal regime is robust provided $h_z \gg J \gg J_\mathrm{U(1)}$.

This U(1)-symmetric prethermal regime allows for a rich dynamical phase diagram featuring a CTC phase~\cite{Bull2022}. The CTC arises under unitary dynamics generated by the effective Hamiltonian~(\ref{eq:XXZ}), hence it is distinct from the more common time crystal realizations in a Floquet setting~\cite{Choi2017,Zhang2017,Randall2021,Mi2022,Kyprianidis2021,Beatrez2023}, where the couplings of the model are externally modulated and the system breaks a discrete symmetry of the drive~\cite{Khemani2016,Else2016,Keyserlingk2016,Yao2017,Ho2017} (for recent reviews of discrete time crystals, see Refs.~\cite{SachaTCReview,KhemaniTCReview,Else2019Review}). While in the stationary frame the CTC occurs at a high energy density, in the rotating frame it corresponds to a near-ground state of the effective Hamiltonian~(\ref{eq:XXZ}). Being at a low temperature in the rotating frame, the CTC can develop an order parameter that spontaneously breaks a symmetry. Whether or not SSB occurs in the rotating frame is determined by the range of interactions $\alpha$.

As discussed above, the ground state of the effective long-range XXZ model~(\ref{eq:XXZ}) exhibits SSB when interactions are sufficiently long-ranged. Dynamically, this manifests as states exhibiting persistent symmetry breaking up to late times. Since the long-range XXZ model is nonintegrable and expected to obey the ETH, whether a particular initial state displays symmetry breaking depends on whether the corresponding finite-temperature Gibbs state also exhibits SSB. For the latter to happen, the couplings $J_x$ and $J_y$ do not necessarily need to be equal to each other, as long as the conditions of the prethermal theorem above hold. What does this imply for the full model~(\ref{eq:Ham})? In the limit $h_z \gg J_\mathrm{U(1)}$, the $h_z$ field effectively drives rotations in the $x-y$ plane, generating an effective U(1) symmetry. However, the underlying anisotropy in the couplings makes rotations in the $x-y$ plane asymmetric~\cite{Bull2022}, leading to small oscillations in the magnitude of the order parameter of the CTC phase $\langle{\sigma^+}\rangle$, which are characteristic of a CTC---see Fig.~\ref{fig:XYZmodel}(b). On the other hand, when interactions are short-ranged and $\alpha$ is above the SSB transition, the order parameter rapidly decays to $\langle{\sigma^+}\rangle = 0$ and now the spectrum respects U(1) symmetry, potentially allowing the QME to emerge when considering two different initial states. Before testing this picture in the numerics, we first introduce the definition of the QME used in this work.

\subsection{Entanglement asymmetry as a probe of the QME}\label{sec:entasym}

The conventional QME setup considers the reduced density matrix $\hat \rho_A \equiv \mathrm{Tr}_{\bar A}|\psi\rangle\langle\psi|$ over a subset of $A$ spins of a pure state $\ket{\psi}$, where $\mathrm{Tr}_{\bar A}$ denotes the partial trace over the complement $\bar A$. Crucially, the chosen state $\ket{\psi}$ is \emph{not} invariant under some global symmetry generated by an operator $\hat Q$. Hence, the symmetry generator restricted to the subsystem, $\hat Q_A$, does not commute with $\hat \rho_A$:
\begin{equation}\label{eq:initialstateQME}
    |\psi\rangle \quad \text{obeys} \quad [ \hat \rho_A, \hat Q_A]\neq 0.
\end{equation}
For simplicity, we will assume $\hat Q=\sum_i \hat \sigma^z_i$ is the generator of the global U(1) symmetry, such that $\hat Q_A=\sum_{i\in A} \hat \sigma^z_i$. 

To quantify the amount of symmetry breaking in $\hat \rho_A$, Ref.~\cite{Ares2023} introduced a quantity called the \emph{entanglement asymmetry},
\begin{equation} \label{eq:S_A}
\Delta S_A \equiv S(\hat \rho_{A,Q})-S(\hat \rho_A), \;\;\; \hat \rho_{A,Q} \equiv \sum_{q\in \mathbb{Z}} \hat\Pi_q \hat \rho_A \hat \Pi_q,
\end{equation}
where $\hat \Pi_q$ is the projector onto the eigenspace of $\hat Q_A$ with charge $q \in \mathbb{Z}$ and $S(\hat\rho) \equiv -\mathrm{Tr}\, \hat\rho \ln \hat\rho$ is the von Neumann entropy. Generally, we have $\Delta S_A \geq 0$ and the equality is attained if and only if  $[ \hat \rho_A,\hat Q_A]=0$. Intuitively,  $\Delta S_A$ quantifies the norm of the off-diagonal blocks connecting different sectors of $\hat Q_A$ in the reduced density matrix $\hat \rho_A$.

Using $\Delta S_A$, we are now in a position to define the QME. We will restrict to cases where we prepare the system in translationally invariant states obeying Eq.~(\ref{eq:initialstateQME}) and parametrized by one or more variables $\theta$ (in the simplest case, $\theta$ may be the angle of each spin on the Bloch sphere). We will say that the QME occurs if the following two conditions are met:
\begin{eqnarray}\label{eq:QME}
    \Delta S_A^{\theta_1}(t=0) &>& \Delta S_A^{\theta_2}(t=0), \\ \label{eq:QME2}
    \Delta S_A^{\theta_1}(t\rightarrow \infty) &<& \Delta S_A^{\theta_2}(t\rightarrow \infty).
\end{eqnarray}
These conditions imply that there is a crossing of the curves $\Delta S_A(t)$ for the two initial states, $\theta_1$ and $\theta_2$. The first crossing $\Delta S_A^{\theta_1}(t{=}\tau_M)=\Delta S_A^{\theta_2}(t{=}\tau_M)$ defines the \emph{Mpemba time}, denoted by $\tau_M$. In the numerical simulations below, it will be convenient to determine $\tau_M$ by examining when the ratio of entanglement asymmetries for two states, $\Delta S_A^{\theta_1}/\Delta S_A^{\theta_2}$, drops below unity.

A few comments are in order. First, in contrast to an open quantum system setting~\cite{Lu2017}, in our case the entanglement asymmetry $\Delta S_A$ is not necessarily a monotonic function of time. Therefore the asymmetry of two states may exhibit multiple crossings during the evolution. In the cases studied in this paper, we have explicitly verified that the results remain qualitatively unchanged if we define the Mpemba time using the \emph{last} instead of the first crossing. More precisely, while the last crossing can result in a larger Mpemba time, it does not have a strong impact on the interplay between QME and SSB. Second, beyond entanglement asymmetry, we note that there are a number of possible alternative definitions of the QME based on information-theoretic measures of how far a system is from equilibrium, instead of how well the state of the system respects a global symmetry~\cite{summer2025}. These measures, such as trace distance~\cite{SM} and the Frobenius norm, allow one to generalize the notion of the QME to systems without global symmetries~\cite{bhore2025, ares2025, ares2025simplerprobeqme, sreejith2025, parez2025, yamashika2025_2}. Since our focus is on symmetry breaking and restoration, we adopt the conventional measure, Eq.~(\ref{eq:S_A}), as a QME diagnostic.

In summary, in the presence of a strong magnetic field and for small differences in transverse spin-spin couplings, the XYZ spin chain is in the robust prethermal regime described by the effective U(1)-symmetric Hamiltonian in Eq.~(\ref{eq:XXZ}). This Hamiltonian undergoes a U(1) SSB transition as the interactions become increasingly long-ranged~\cite{Maghrebi2017, Gong2016Kaleidoscope, Gong2016TopologicalPhases}. We conjecture that the occurrence of the QME in this model depends on two conditions: (i) the robustness of the effective XXZ model as the field strength, $h_z$, is increased, and (ii) the emergence of SSB in the equilibrium XXZ model as the interaction length scale, $\alpha^{-1}$, is increased. In the following sections, we test this conjecture in numerical simulations. We will detect the QME using the entanglement asymmetry, $\Delta S_A$, for different initial states, where their intersection is identified as the QME. 

\begin{figure*}
    \centering
    \includegraphics[width=\linewidth]{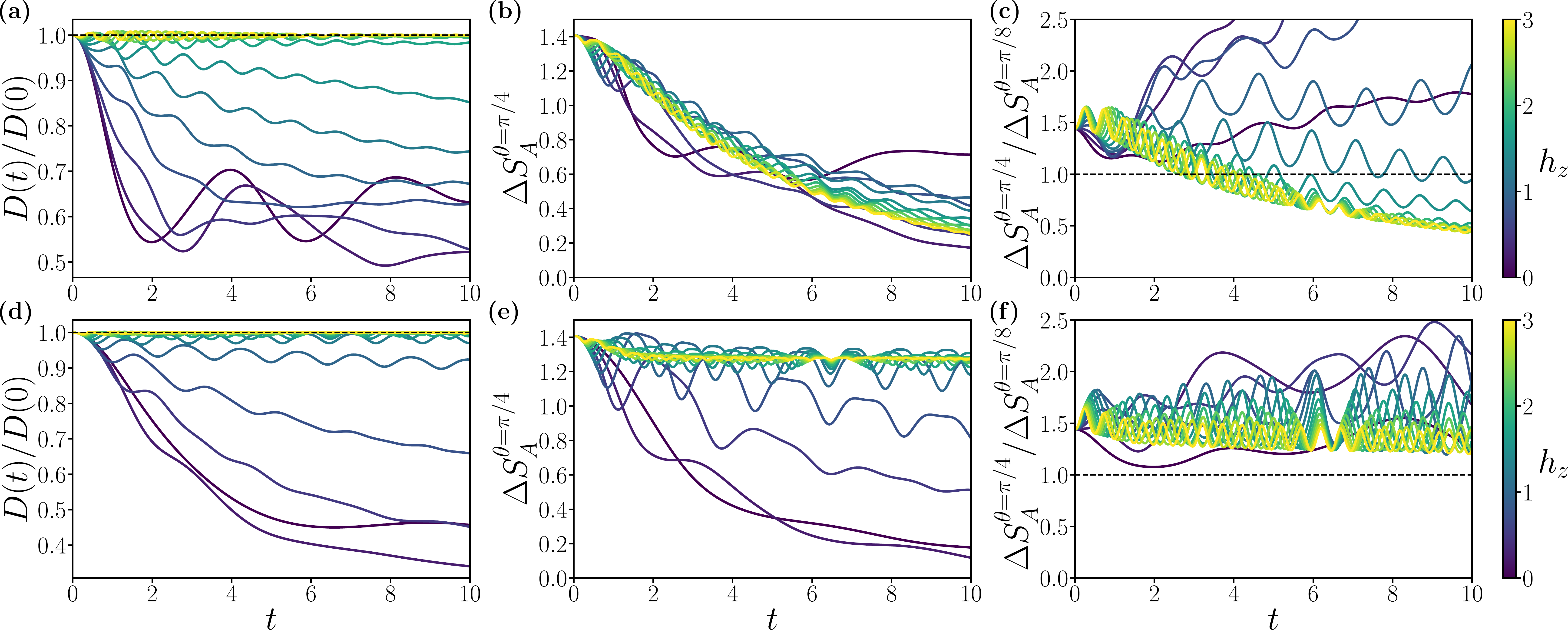}
    \caption{Quench dynamics for the initial states in Eq.~(\ref{eq:Initial_state}) under the long-range XYZ Hamiltonian in Eq.~(\ref{eq:Ham}). The parameters are $J_x=-0.5$, $J_y=-1.5$ and $J_z=-0.75$, at $\alpha=4$ (top) and $1.5$ (bottom), while the color bar shows the field strength $h_z$. (a) and (b) show the expectation value $D(t) \equiv \langle \psi(t)|\hat D|\psi(t)\rangle$ of the U(1)-invariant effective XXZ Hamiltonian in Eq.~(\ref{eq:D}) for the $\theta=\pi/4$ initial state at $\alpha=4$ (a) and $1.5$ (b). We normalize the curves by $D(t=0)$. The prethermal regime, $D(t)/D(0)\approx 1$, is seen to be robust for both short- and long-range interactions if the field strength is sufficiently large ($h_z\gtrsim 2$). (c) and (d) show the corresponding entanglement asymmetry $\Delta S_A$ for $N_A=4$ sites for $\alpha=4$ (c) and $1.5$ (d). The symmetry is restored for short-range interactions (c), but not for long-range (d). (e) and (f) show the ratio of the entanglement asymmetry at $\theta_1=\pi/4$ and $\theta_2=\pi/8$ at $\alpha=4$ (e) and $1.5$ (f). The drop of the large-$h_z$ (red) curves below unity in (e) signals the QME, which is absent for longer-range interactions in (f). All data were obtained for a subsystem of $N_A=4$ spins in an infinite chain using TDVP with translation-invariant iMPS of bond dimension $\chi=128$ and a timestep $\delta t=0.025$.}
    \label{fig:XYZ_results}
\end{figure*}

\section{Varying the field strength}\label{sec:resXYZ}

We now perform a numerical analysis of the dynamical phase diagram of the model~(\ref{eq:Ham}) by varying the field strength $h_z$. We consider quenches from uniformly tilted product initial states:
\begin{equation} \label{eq:Initial_state}
\ket{\psi(t=0,\theta)}=\bigotimes_j \exp\left(i\theta \hat \sigma^y_j\right)\ket{\uparrow}_j,
\end{equation}
which break the U(1) symmetry (magnetization in the $z$ direction), as prescribed in Eq.~(\ref{eq:initialstateQME}). We will work in the infinite system-size limit, simulating the dynamics using the time-dependent variational principle (TDVP) for translation-invariant infinite matrix-product states (iMPS)~\cite{Haegeman,Haegeman2016,Vanderstraeten2019},
approximating the long-range couplings as a sum of exponential
terms~\cite{Pirvu2010} (see SM~\cite{SM} for details of our implementation of long-range interactions). Working directly in the thermodynamic limit greatly simplifies the numerical analysis of SSB, since the dynamics in finite systems exhibit $O(N^{1/2})$ system-size-dependent oscillations due to the splitting in the ground-state manifold~\cite{Anderson1952,Tasaki2019} (see SM~\cite{SM} for a system-size-scaling analysis). For the family of states in Eq.~(\ref{eq:Initial_state}), the initial entanglement asymmetry $\Delta S_A(t=0)$ monotonically increases as $\theta$ varies from $\theta=0$, where $\ket{\psi(t=0)}=\ket{\uparrow\uparrow\uparrow\ldots}$, to $\theta=\pi/4$, where $\ket{\psi(t=0)}=\ket{+++\ldots}$. 

In Fig.~\ref{fig:XYZ_results} we present quench dynamics for the initial states defined in Eq.~(\ref{eq:Initial_state}) and evolved under the long-range XYZ Hamiltonian in Eq.~(\ref{eq:Ham}) with $J_x=-0.5$, $J_y=-1.5$ and $J_z=-0.75$, at $\alpha=4$ (upper panels) and $\alpha=1.5$ (lower panels) for increasing field strength $h_z$. Figures~\ref{fig:XYZ_results}(a) and~\ref{fig:XYZ_results}(b) show the expectation value of the prethermal Hamiltonian~(\ref{eq:D}), $D(t) \equiv \langle \psi(t)|\hat D|\psi(t)\rangle$ (normalized by its value at $t=0$) for the $\theta=\pi/4$ state at $\alpha=4$ (a) and  $1.5$ (b). For small $h_z$ the system rapidly leaves the prethermal regime, as indicated by the sharp drop of $D(t)/D(0)$ below one. In contrast, at large $h_z$, $D(t)/D(0)$ remains close to one even at late times, demonstrating that the prethermal Hamiltonian accurately captures the dynamics. There are small but noticeable oscillations in $D(t)/D(0)$ as $h_z$ increases. As explained in Sec.~\ref{sec:CTC}, these arise because the field drives rotations in the $x$-$y$ plane, while the asymmetry in the spin-spin couplings in the $x$ and $y$ directions leads to slightly elliptic rotations. These small oscillations are characteristic of the prethermal CTC~\cite{Bull2022}. It is important to note that this oscillatory behavior is due to subleading corrections to the exact U(1)-symmetric prethermal Hamiltonian in Eq.~(\ref{eq:XXZ}). 

Figures~\ref{fig:XYZ_results}(c) and~\ref{fig:XYZ_results}(d) show the evolution of the entanglement asymmetry $\Delta S_A$ for the $\theta=\pi/4$ initial state at  $\alpha=4$ (c) and $1.5$ (d). At large $h_z$, the prethermal phase is robust and there is an emergent U(1) symmetry. As discussed above, when $\alpha$ is large the system is in the U(1)-preserving XY phase, so we expect to observe symmetry restoration. This can be seen in Fig.~\ref{fig:XYZ_results}(c) where the entanglement asymmetry tends to $0$ for large $h_z$. In contrast, for $\alpha=1.5$ the system is in the U(1)-breaking superfluid phase, hence in Fig.~\ref{fig:XYZ_results}(d) the entanglement asymmetry remains constant at late times for large enough $h_z$. 

Altogether, Figs.~\ref{fig:XYZ_results}(a)-\ref{fig:XYZ_results}(d) serve as a diagnostic of the system being in the prethermal regime, confirming that the quench dynamics matches that of the effective Hamiltonian~(\ref{eq:D}) at large $h_z$. To directly reveal the QME, in Figs.~\ref{fig:XYZ_results}(e) and~\ref{fig:XYZ_results}(f) we show the ratio of entanglement asymmetries for two initial states of the form in Eq.~(\ref{eq:Initial_state}) with $\theta_1=\pi/4$ and $\theta_2=\pi/8$, plotted
as a function of time for $\alpha=4$ and $1.5$,  respectively. For our choice of parameters, the asymmetry ratio starts out above 1 at $t=0$, and the QME occurs if it drops below 1 at some later time due to the crossing of the $\Delta S_A$ curves traced by the two initial states. At small $h_z$, for either value of $\alpha$, the system manifestly breaks U(1) symmetry and, unsurprisingly, we do not observe the ratio dropping below unity, consistent with an absence of the QME. In contrast, at large $h_z$ the emergent U(1) symmetry leads to strikingly different behavior for $\alpha=4$ compared to $\alpha=1.5$. For $\alpha=4$, the U(1)-preserving behavior of the equilibrium physics leads to symmetry restoration in the nonequilibrium dynamics, subsequently leading to the QME in Fig.~\ref{fig:XYZ_results}(e). As the field strength decreases and the prethermal timescale of the model becomes shorter, the Mpemba time sharply diverges. For $\alpha=1.5$, the SSB in the equilibrium model prevents symmetry restoration in the non-equilibrium dynamics, thereby causing the QME to cease, as shown in Fig.~\ref{fig:XYZ_results}(f).

To summarize, Fig.~\ref{fig:XYZ_results} demonstrates the correlation between the entanglement asymmetry and the existence of the prethermal Hamiltonian. When $h_z$ is small, the prethermal phase is not robust; therefore there is no emergent U(1) symmetry and dynamical symmetry restoration does not occur, resulting in the absence of a QME. However, when $h_z$ is sufficiently large, the system remains in the prethermal phase where the prethermal XXZ description is valid (due to the prethermalization timescale being sufficiently long). In this regime, symmetry restoration follows the dynamics expected under the prethermal Hamiltonian, where large $\alpha$ implies that the system is in the U(1)-conserving phase leading to dynamical symmetry restoration and the QME, while small $\alpha$ leads to SSB and no symmetry restoration, halting the QME. Thus by tuning $h_z$ and $\alpha$, the presence of the QME for the class of states in Eq.~(\ref{eq:Initial_state}) can be controlled by steering toward or away from symmetry restoration.

\section{Tuning the interaction range and energy of initial states}\label{sec:resXXZ}

In the previous section, we showed that as $h_z$ increased, the dynamics of the long-range XYZ model in the rotating frame is governed by an effective long-range XXZ model, with qualitatively different behavior for the entanglement asymmetry and QME depending on the value of $\alpha$. In this section, we work in the $h_z\to\infty$ limit and explicitly study the long-range XXZ model to ascertain the dependence of the QME on $\alpha$ and $J_z$. We will discuss how the QME depends upon the choice of initial state, in particular the energy density. 

\begin{figure}[tb!]
    \centering
    \includegraphics[width=1\linewidth]{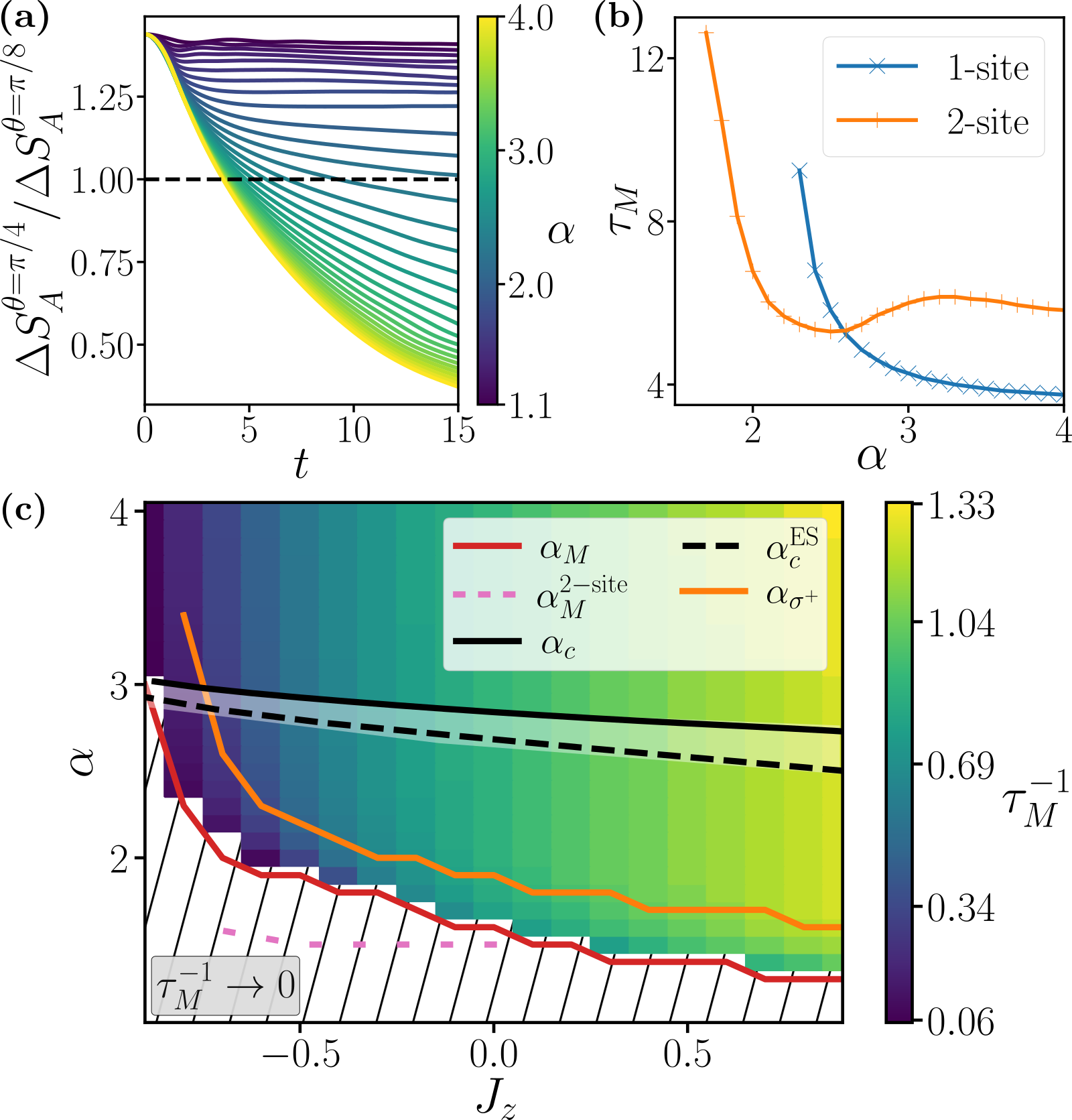}
    \caption{(a) Ratio of the entanglement asymmetry for the initial states in Eq.~(\ref{eq:Initial_state}) with $\theta_1=\pi/4$ and $\theta_2=\pi/8$, evolved under the XXZ Hamiltonian~(\ref{eq:D}). Parameters are $J_x=J_y=-1$ and $J_z=-0.75$ with varying $\alpha$ (color bar). (b) The Mpemba time $\tau_M$ as a function of $\alpha$ for these states (labeled $1$-site), as well as for two tilted N\'eel states in Eq.~(\ref{eq:high_energy_state}) with $\phi_1=\pi/4$ and $\phi_2=\pi/8$ (labeled $2$-site), evolved under the same Hamiltonian. (c) The inverse of the Mpemba time $\tau^{-1}_M$ (color bar) as a function of $(J_z,\alpha)$. The red line labeled $\alpha_M$ marks the region where $\tau_M\rightarrow\infty$ and no QME occurs for the tilted ferromagnetic states in Eq.~(\ref{eq:Initial_state}), while the pink dotted line  $\alpha_M^{2\mathrm{-site}}$ denotes the same for the tilted N\'eel states in Eq.~(\ref{eq:high_energy_state}) (this line was only computed in the interval $J_z\in[-0.75,0]$). The SSB phase boundary $\alpha_c$ is defined by central charge $c_\mathrm{eff}$ deviating from unity by more than $10\%$. $c_\mathrm{eff}$ was calculated from finite DMRG comparing the difference in entanglement entropy between system sizes $N=192$ and $200$, following the method in Refs.~\cite{Maghrebi2017, Gong2016Kaleidoscope, Gong2016TopologicalPhases}, at a maximum bond dimension $\chi=250$. An alternative estimate of the SSB boundary, $\alpha^{\mathrm{ES}}_c$, was obtained following Ref.~\cite{Schneider2022}, where the entanglement spectrum of the ground state was calculated for a system of $N=192$ spins using DMRG with $\chi=250$. The shaded area between $\alpha_c$ and $\alpha_c^\mathrm{ES}$ represents the uncertainty of the SSB boundary. The line $\alpha_{\sigma^+}$ was obtained by determining the $\alpha$ at which the ratio of the order parameters $\langle\sigma^+\rangle^{\theta=\pi/4} / \langle\sigma^+\rangle^{\theta=\pi/8}$ stays above $1$. All dynamics data were obtained for a subsystem of $N_A=4$ spins using TDVP for iMPS with time step $\delta t=0.025$ and maximum bond dimension $\chi=128$. }
    \label{fig:XXZ_results}
\end{figure}

Using the tilted product initial states previously introduced in Eq.~(\ref{eq:Initial_state}), we analyze the quench dynamics under the XXZ Hamiltonian in Eq.~(\ref{eq:D}) with $J_x=J_y=-1$ and varying $\alpha$ and $J_z$. Figure~\ref{fig:XXZ_results}(a) shows the entanglement asymmetry ratio for $\theta_1 =\pi/4$ and $\theta_2=\pi/8$  with fixed $J_z=-0.75$ and varying $\alpha$. Consistent with Figs.~\ref{fig:XYZ_results}(e) and~\ref{fig:XYZ_results}(f), we notice that increasing the range of interactions halts the QME. Since the long-range XXZ model is nonintegrable, these states are all expected to relax to thermal equilibrium. Therefore the halting of the QME as $\alpha$ is decreased depends on whether the thermal state exhibits SSB or not. This can be quantified using the normalized energy density 
\begin{equation}
 \epsilon=(E-E_{\min})/(E_{\max}-E_{\min}),   
\end{equation}
where $E$ is the energy of the state, and $E_{\min}$, $E_{\max}$ are those of the lowest and highest energy states, respectively. Larger values of $\epsilon$ equilibrate to higher temperature thermal states, and consequently require longer interaction ranges to stabilize the U(1) SSB. 

The initial states in Eq.~(\ref{eq:Initial_state}) are low-energy density states, e.g., we have $\epsilon=0.003$ for $\theta=\pi/4$ and $\epsilon=0.068$ for $\theta=\pi/8$ at $J_z=-0.75$ and $\alpha=2$.  However, SSB is expected to occur at smaller $\alpha$ if we consider even higher energy density states. To probe this and access a broader range of $\epsilon$, we consider a class of tilted N\'eel states:
\begin{equation}
    \ket{\varphi(t=0,\phi)} = \bigotimes_i \Big(\ket{\uparrow}_{2i-1}\otimes[\cos{\phi}\ket{\uparrow}_{2i} - \sin{\phi}\ket{\downarrow}_{2i}] \Big). \label{eq:high_energy_state}
\end{equation}
Figure~\ref{fig:XXZ_results}(b) shows how the Mpemba time $\tau_M$ changes as a function of $\alpha$ for the one-site periodic tilted product intial states in Eq.~(\ref{eq:Initial_state}) with $\theta_1= \pi/4$, $\theta_2=\pi/8$ (labeled $1$-site), and the two-site periodic tilted N\'eel states in Eq.~(\ref{eq:high_energy_state}) with $\phi_1= \pi/4$, $\phi_2=\pi/8$ (labeled $2$-site). For the tilted N\'eel states the normalized energy densities are $\epsilon=0.407$ for $\phi=\pi/4$ and $\epsilon=0.209$ for $\phi=\pi/8$  at $J_z=-0.75$ and $\alpha=2$~\cite{SM}, placing them significantly higher in the spectrum compared to the tilted ferromagnetic states. We find that the Mpemba time for the tilted product states diverges and halts at a shorter range of interactions than for the tilted N\'eel states.   This is due to the fact that the critical $\alpha$ required to stabilize the U(1) SSB phase decreases with increasing energy density, as the tilted N\'eel states are expected to thermalize to higher-temperature Gibbs states compared to the tilted ferromagnetic states.

In Fig.~\ref{fig:XXZ_results}(c), the extracted Mpemba time $\tau_M$ is plotted for a range of $(J_z,\alpha)$ values. The critical line, denoted by $\alpha_M$, represents the largest value of $\alpha$ for which we do not observe the QME for the tilted ferromagnetic states in Eq.~(\ref{eq:Initial_state}). Below this line is a region in which the QME does not occur ($\tau_M\rightarrow\infty$). The determined $\alpha_M$, in principle, depends on the maximum evolution time; however, the plotted value is converged for all $J_z$ values except in the immediate vicinity of $J_z=-1$~\cite{SM}. We also calculate $\alpha_M^{2\mathrm{-site}}$, the $\alpha$ at which the QME halts for the tilted N\'eel states in Eq.~(\ref{eq:high_energy_state}). We find that it lies below the $\alpha_M$ found for tilted ferromagnetic states, consistent with Fig.~\ref{fig:XXZ_results}(b).

This demonstrates that the QME boundary depends on the choice of initial states, with their energy density providing an important tunable parameter. The onset of the QME can also be probed by the order parameter $\langle \sigma^+ \rangle$, as shown in Fig.~\ref{fig:XYZmodel}(b). Motivated by this, we compute the ratio of $\langle \sigma^+ \rangle$ for the $\theta=\pi/4$ and $\theta=\pi/8$ tilted ferromagnetic states, and find that it closely mirrors the corresponding ratio of entanglement asymmetries. This gives us another QME boundary estimate, denoted $\alpha_{\sigma^+}$ in Fig.~\ref{fig:XXZ_results}(c), which serves as an upper bound for $\alpha_M$. In the special case of subsystem size $N_A=1$, the two are directly related since the off-diagonal elements of $\hat\rho_A$ are proportional to $\langle\sigma^+\rangle$.

An immediate question is how does the $\alpha_M$ critical line compare with the SSB phase boundary? To compare the onset of SSB with the $\alpha_M$ line, we estimated the former using two complementary methods. We determine the location of SSB---denoted by $\alpha_c$ in Fig.~\ref{fig:XXZ_results}(c)---following the method in Refs.~\cite{Maghrebi2017, Gong2016Kaleidoscope, Gong2016TopologicalPhases} based on the central charge $c_\mathrm{eff}$ extracted from entanglement entropy scaling in the ground state at a conformally invariant 1D critical point~\cite{Calabrese2004, Calabrese2009}, obtained from
finite-size DMRG using the ITensor
library~\cite{ITensor, fishman2025itensor-r0.3}. At large values of $\alpha$,  $c_\mathrm{eff}=1$ is quantized due to conformal symmetry~\cite{Alcaraz1992, Ejima2015}; as we reduce $\alpha$, $c_\mathrm{eff}$ serves as a diagnostic of whether the model has left the XY phase. Eventually, in the SSB phase, $c_\mathrm{eff}$ is not constrained by conformal symmetry~\cite{Maghrebi2017, Vodola2014, Vodola2015} and can generally deviate from unity. In practice, we find that the SSB phase has a varying $c_\mathrm{eff}$ that is greater than 1.  We therefore define the SSB point by $c_\mathrm{eff}$ deviating from 1 by more than $10\%$. 
As a consistency check, we have used a different method for estimating the SSB point based on the entanglement spectrum~\cite{Schneider2022}. This method produces an alternative estimate of the SSB phase boundary, labeled $\alpha^{\mathrm{ES}}_c$ in Fig.~\ref{fig:XXZ_results}(c). The two estimates, $\alpha_c$ and $\alpha_c^\mathrm{ES}$, are consistent with each other, and the small quantitative differences between them can be viewed as the uncertainty in determining the SSB phase boundary.

Generally, from Fig.~\ref{fig:XXZ_results}(c) we infer that $\alpha_M < \alpha_c$; thus U(1) SSB serves as an upper bound for the occurrence of the QME. The physical origin of this separation is that the two boundaries probe different notions of order: $\alpha_c$ is a ground-state property and determines SSB  at zero temperature, whereas $\alpha_M$ is a dynamical, state-dependent boundary extracted from finite-energy quenches. If the system thermalizes, the relevant states are characterized by an effective temperature, where SSB  is generally less robust than in the ground state. Consequently, longer-range interactions (smaller $\alpha$) are needed to halt dynamical symmetry restoration, leading naturally to $\alpha_M < \alpha_c$.

This leaves an interesting intermediate regime $\alpha_M < \alpha < \alpha_c$, where the ground state is already ordered, yet the excited states accessed dynamically still permit a finite Mpemba time. Accordingly, the halting of the QME depends on the energy density of the initial states: as the energy density decreases, the line $\alpha_M$ shifts closer to the SSB phase boundary, while at higher energies the two boundaries move further apart, as confirmed in Fig.~\ref{fig:XXZ_results}(b). 
Even for low-energy product states, a residual separation between $\alpha_M$ and $\alpha_c$ can remain because low energy density does not by itself imply proximity to the correlated ground state. In particular, the product states used here lack the correlations present in the ordered phase, which must first develop dynamically after the quench. Thus, the buildup of SSB order can occur on longer timescales than the Mpemba crossing, allowing a finite $\tau_M$ even where the ground state is already ordered. 

Finally, the discussion above holds away from $J_z=-1$. As $J_z$ approaches $J_z=-1$, a finite Mpemba time cannot be detected within accessible timescales. This is due to the SU(2) symmetry of the $J_z=-1$ point, with the tilted ferromagnetic states of Eq.~(\ref{eq:Initial_state}) all becoming exact eigenstates of the Hamiltonian. From this we can infer that as $J_z\rightarrow-1$, $\tau_M$ will also diverge for $\alpha > \alpha_c$ due to the choice of initial states. Hence, in this limit, the SSB phase boundary no longer bounds the region where the QME will halt, with the QME halting even in the U(1) conserving region for the tilted ferromagnetic states. In SM~\cite{SM}, we analyze the SU(2) point more closely, finding that the QME with respect to the SU(2) symmetry does not halt, which we attribute to the lack of SU(2) SSB in the ground state of the model.

\section{Discussion and conclusions}\label{sec:conc}

In this paper, we explored the interplay between the QME and SSB by varying the range of interactions. We analyzed both the full XYZ model and its $h_z \to \infty$ limit, where the dynamics reduce effectively to the XXZ model. We showed that tuning the distance from this limit allows one to control the QME, while in the XXZ limit, SSB can be exploited to suppress the QME by adjusting the interaction range. Finally, we demonstrated that the effective temperature of the initial state plays a crucial role, as the interaction range at which SSB emerges depends sensitively on the state’s energy density.

\begin{figure}[tb]
    \centering
    \includegraphics[width=\linewidth]{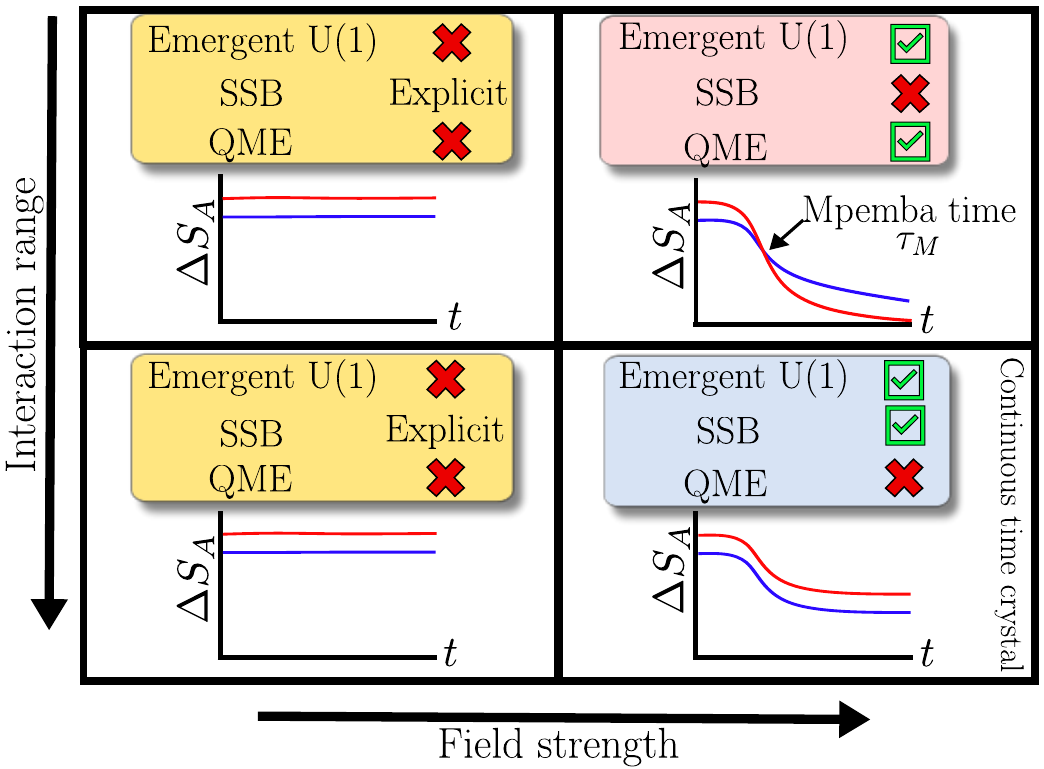}
    \caption{Dynamical phase diagram of U(1) symmetry restoration in the XYZ model~(\ref{eq:Ham}) with generic couplings that break the U(1) symmetry, as a function of the field strength $h_z$ and interaction range $\alpha$. The QME is only possible when the field strength is large, such that the spin couplings become isotropic in the $x$-$y$ plane, and the U(1) symmetry is preserved because the interactions are effectively short-ranged. We note that this phase diagram is expected to be valid over the prethermal timescale~\cite{Else2017}, Eq.~(\ref{eq:XXZ}), which depends exponentially on $h_z$ and can be made much longer than the Mpemba time $\tau_M$. Moreover, the phase diagram is, in principle, dependent on the types of initial states, which we assume here to be product states of spins pointing in the same direction on the Bloch sphere [Eq.~(\ref{eq:Initial_state})] or tilted N\'eel states [Eq.~(\ref{eq:high_energy_state})], or their combinations (see text for details).
    }
    \label{fig:sketch}
\end{figure}

Combining the results of Sec.~\ref{sec:resXYZ} and~\ref{sec:resXXZ}, we arrive at a dynamical symmetry-restoration phase diagram in Fig.~\ref{fig:sketch}. The QME occurs in the top right corner, corresponding to large fields and sufficiently short-range interactions. This diagram applies to the XYZ model with generally unequal spin–spin couplings in the $x$ and $y$ directions, hence without an explicit U(1) symmetry. A potential caveat of Fig.~\ref{fig:sketch} is its dependence on the choice of initial states. This phase diagram is consistent with our results for the tilted product states~(\ref{eq:Initial_state}) and the tilted N\'eel states~(\ref{eq:high_energy_state}), while in the SM~\cite{SM} we demonstrate that similar behavior also holds for random MPS states with bond dimension $\chi=2$. Thus we expect that a broad class of experimentally preparable initial states follows the dynamical behavior in Fig.~\ref{fig:sketch}. Nevertheless, to clearly observe the effect, the initial entanglement-asymmetry ratio must be sufficiently larger than 1.

We note that signatures of the QME have recently been observed in small-scale trapped-ion experiments, which realize the long-range XY spin model~\cite{Joshi2024}. However, Ref.~\cite{Joshi2024} did not explore the effect of SSB or more generic couplings that make the model interacting and break U(1) symmetry. A theoretical investigation of the QME in long-range spin systems was also carried out in Ref.~\cite{yamashika2025} using time-dependent spin-wave theory, which demonstrated that QME behavior can arise in the semiclassical regime. That work primarily considers finite-size systems and focuses on very long-range interactions ($\alpha<1$), where symmetry restoration is driven by quantum fluctuations of a precessing ferromagnetic state.  Our results address a distinct regime: we work directly in the thermodynamic limit and over a broader range of interaction exponents, where SSB can qualitatively modify the long-time dynamics. In this limit, symmetry need not be restored, and the existence of the QME is sharply tied to whether the system lies in a symmetry-restoring or symmetry-broken phase. This interplay between SSB and dynamical symmetry restoration is essential for understanding when the QME can be suppressed or halted in generic long-range systems with anisotropic interactions.

In this work, we found that an emergent U(1) symmetry could be established using a strong longitudinal field in a Hamiltonian system. Alternatively, U(1) symmetry could arise from Floquet driving, which should lead to analogous behavior for the QME, similar to the prethermal Floquet time cystal scenario~\cite{Machado2020,Kyprianidis2021,Beatrez2023}. Moreover, it would be interesting to study other models that realize SSB of a non-Abelian symmetry. Beyond 1D, the HMW theorem also applies to short-range interacting systems in $d=2$ spatial dimensions, but can be evaded using sufficiently long-range algebraic interactions. Therefore we expect a similar halting of the QME in models such as the long-range spin-$1/2$ Heisenberg model in $d=2$ as the length scale of interactions is tuned~\cite{Zhao_2023}. By contrast, in $d=3$ SSB becomes possible even with short-range interactions, hence it would be particularly interesting to explore the interplay with the QME in this case. While this is challenging for classical numerics, it would be a suitable task for quantum simulations as the relatively short Mpemba time makes the problem amenable to the current generation of noisy quantum hardware. 

\begin{acknowledgments}

We would like to thank Tanmay Bhore and Jie Ren for useful discussions. We acknowledge the use of the ITensor
library~\cite{ITensor, fishman2025itensor-r0.3} for the DMRG calculations
presented in this work. We acknowledge support by the Leverhulme Trust Research Leadership Award No. RL-2019-015 and EPSRC Grants EP/Z533634/1,  EP/W524372/1, UKRI1337, UKRI3780.
This research was supported in part by grant NSF PHY-2309135 to the Kavli Institute for Theoretical Physics (KITP). Z.P. acknowledges support by the Erwin
Schrödinger International Institute for Mathematics and Physics. I.M.  acknowledges support by the US Department of Energy, Office of
Science, Basic Energy Sciences, Materials Sciences and
Engineering Division.

\end{acknowledgments}

\section*{Data Availability}
The data that support the findings of
this article are openly available \cite{dataset2026}.

\bibliography{references}

@Misc{SM,
  Note                     = {See Supplemental Material for further details and numerical results.
}
}

@article{Lu2017,
author = {Zhiyue Lu  and Oren Raz },
title = {Nonequilibrium thermodynamics of the Markovian Mpemba effect and its inverse},
journal = {Proceedings of the National Academy of Sciences},
volume = {114},
number = {20},
pages = {5083-5088},
year = {2017},
doi = {10.1073/pnas.1701264114},
URL = {https://www.pnas.org/doi/abs/10.1073/pnas.1701264114}}

@article{Ho2017,
  title = {Critical Time Crystals in Dipolar Systems},
  author = {Ho, Wen Wei and Choi, Soonwon and Lukin, Mikhail D. and Abanin, Dmitry A.},
  journal = {Phys. Rev. Lett.},
  volume = {119},
  issue = {1},
  pages = {010602},
  numpages = {6},
  year = {2017},
  month = {Jul},
  publisher = {American Physical Society},
  doi = {10.1103/PhysRevLett.119.010602},
  url = {https://link.aps.org/doi/10.1103/PhysRevLett.119.010602}
}

@Article{Choi2017,
author={Choi, Soonwon
and Choi, Joonhee
and Landig, Renate
and Kucsko, Georg
and Zhou, Hengyun
and Isoya, Junichi
and Jelezko, Fedor
and Onoda, Shinobu
and Sumiya, Hitoshi
and Khemani, Vedika
and von Keyserlingk, Curt
and Yao, Norman Y.
and Demler, Eugene
and Lukin, Mikhail D.},
title={Observation of discrete time-crystalline order in a disordered dipolar many-body system},
journal={Nature},
year={2017},
month={Mar},
day={01},
volume={543},
number={7644},
pages={221-225},
issn={1476-4687},
doi={10.1038/nature21426},
url={https://doi.org/10.1038/nature21426}
}

@Article{Zhang2017,
author={Zhang, J.
and Hess, P. W.
and Kyprianidis, A.
and Becker, P.
and Lee, A.
and Smith, J.
and Pagano, G.
and Potirniche, I.-D.
and Potter, A. C.
and Vishwanath, A.
and Yao, N. Y.
and Monroe, C.},
title={Observation of a discrete time crystal},
journal={Nature},
year={2017},
month={Mar},
day={01},
volume={543},
number={7644},
pages={217-220},
issn={1476-4687},
doi={10.1038/nature21413},
url={https://doi.org/10.1038/nature21413}
}

@article{
Randall2021,
author = {J. Randall  and C. E. Bradley  and F. V. van der Gronden  and A. Galicia  and M. H. Abobeih  and M. Markham  and D. J. Twitchen  and F. Machado  and N. Y. Yao  and T. H. Taminiau },
title = {Many-body localized discrete time crystal with a programmable spin-based quantum simulator},
journal = {Science},
volume = {374},
number = {6574},
pages = {1474-1478},
year = {2021},
doi = {10.1126/science.abk0603},
URL = {https://www.science.org/doi/abs/10.1126/science.abk0603},
eprint = {}}

@Article{Beatrez2023,
author={Beatrez, William
and Fleckenstein, Christoph
and Pillai, Arjun
and de Leon Sanchez, Erica
and Akkiraju, Amala
and Diaz Alcala, Jesus
and Conti, Sophie
and Reshetikhin, Paul
and Druga, Emanuel
and Bukov, Marin
and Ajoy, Ashok},
title={Critical prethermal discrete time crystal created by two-frequency driving},
journal={Nature Physics},
year={2023},
month={Mar},
day={01},
volume={19},
number={3},
pages={407-413},
issn={1745-2481},
doi={10.1038/s41567-022-01891-7},
url={https://doi.org/10.1038/s41567-022-01891-7}
}

@article{Khemani2016,
  title = {Phase Structure of Driven Quantum Systems},
  author = {Khemani, Vedika and Lazarides, Achilleas and Moessner, Roderich and Sondhi, S. L.},
  journal = {Phys. Rev. Lett.},
  volume = {116},
  issue = {25},
  pages = {250401},
  numpages = {6},
  year = {2016},
  month = {Jun},
  publisher = {American Physical Society},
  doi = {10.1103/PhysRevLett.116.250401},
  url = {https://link.aps.org/doi/10.1103/PhysRevLett.116.250401}
}

@article{Else2016,
  title = {Floquet Time Crystals},
  author = {Else, Dominic V. and Bauer, Bela and Nayak, Chetan},
  journal = {Phys. Rev. Lett.},
  volume = {117},
  issue = {9},
  pages = {090402},
  numpages = {5},
  year = {2016},
  month = {Aug},
  publisher = {American Physical Society},
  doi = {10.1103/PhysRevLett.117.090402},
  url = {https://link.aps.org/doi/10.1103/PhysRevLett.117.090402}
}

@article{SachaTCReview,
   title={Time crystals: a review},
   volume={81},
   ISSN={1361-6633},
   url={http://dx.doi.org/10.1088/1361-6633/aa8b38},
   DOI={10.1088/1361-6633/aa8b38},
   number={1},
   journal={Reports on Progress in Physics},
   publisher={IOP Publishing},
   author={Sacha, Krzysztof and Zakrzewski, Jakub},
   year={2017},
   month={Nov},
   pages={016401}
}

@misc{KhemaniTCReview,
  doi = {10.48550/ARXIV.1910.10745},
  
  eprint = {1910.10745},
  
  archivePrefix = {arXiv},
  
  url = {https://arxiv.org/abs/1910.10745},
  
  author = {Khemani, Vedika and Moessner, Roderich and Sondhi, S. L.},
  
  title = {A Brief History of Time Crystals},
  
  publisher = {arXiv},
  
  year = {2019},
  
  copyright = {arXiv.org perpetual, non-exclusive license}
}

@Article{Mi2022,
author={Mi, Xiao
and Ippoliti, Matteo
and Quintana, Chris
and Greene, Ami
and Chen, Zijun
and Gross, Jonathan
and Arute, Frank
and Arya, Kunal
and Atalaya, Juan
and Babbush, Ryan
and Bardin, Joseph C.
and Basso, Joao
and Bengtsson, Andreas
and Bilmes, Alexander
and Bourassa, Alexandre
and Brill, Leon
and Broughton, Michael
and Buckley, Bob B.
and Buell, David A.
and Burkett, Brian
and Bushnell, Nicholas
and Chiaro, Benjamin
and Collins, Roberto
and Courtney, William
and Debroy, Dripto
and Demura, Sean
and Derk, Alan R.
and Dunsworth, Andrew
and Eppens, Daniel
and Erickson, Catherine
and Farhi, Edward
and Fowler, Austin G.
and Foxen, Brooks
and Gidney, Craig
and Giustina, Marissa
and Harrigan, Matthew P.
and Harrington, Sean D.
and Hilton, Jeremy
and Ho, Alan
and Hong, Sabrina
and Huang, Trent
and Huff, Ashley
and Huggins, William J.
and Ioffe, L. B.
and Isakov, Sergei V.
and Iveland, Justin
and Jeffrey, Evan
and Jiang, Zhang
and Jones, Cody
and Kafri, Dvir
and Khattar, Tanuj
and Kim, Seon
and Kitaev, Alexei
and Klimov, Paul V.
and Korotkov, Alexander N.
and Kostritsa, Fedor
and Landhuis, David
and Laptev, Pavel
and Lee, Joonho
and Lee, Kenny
and Locharla, Aditya
and Lucero, Erik
and Martin, Orion
and McClean, Jarrod R.
and McCourt, Trevor
and McEwen, Matt
and Miao, Kevin C.
and Mohseni, Masoud
and Montazeri, Shirin
and Mruczkiewicz, Wojciech
and Naaman, Ofer
and Neeley, Matthew
and Neill, Charles
and Newman, Michael
and Niu, Murphy Yuezhen
and O'Brien, Thomas E.
and Opremcak, Alex
and Ostby, Eric
and Pato, Balint
and Petukhov, Andre
and Rubin, Nicholas C.
and Sank, Daniel
and Satzinger, Kevin J.
and Shvarts, Vladimir
and Su, Yuan
and Strain, Doug
and Szalay, Marco
and Trevithick, Matthew D.
and Villalonga, Benjamin
and White, Theodore
and Yao, Z. Jamie
and Yeh, Ping
and Yoo, Juhwan
and Zalcman, Adam
and Neven, Hartmut
and Boixo, Sergio
and Smelyanskiy, Vadim
and Megrant, Anthony
and Kelly, Julian
and Chen, Yu
and Sondhi, S. L.
and Moessner, Roderich
and Kechedzhi, Kostyantyn
and Khemani, Vedika
and Roushan, Pedram},
title={Time-crystalline eigenstate order on a quantum processor},
journal={Nature},
year={2022},
month={Jan},
day={01},
volume={601},
number={7894},
pages={531-536},
issn={1476-4687},
doi={10.1038/s41586-021-04257-w},
url={https://doi.org/10.1038/s41586-021-04257-w}
}

@article{
Kyprianidis2021,
author = {A. Kyprianidis  and F. Machado  and W. Morong  and P. Becker  and K. S. Collins  and D. V. Else  and L. Feng  and P. W. Hess  and C. Nayak  and G. Pagano  and N. Y. Yao  and C. Monroe },
title = {Observation of a prethermal discrete time crystal},
journal = {Science},
volume = {372},
number = {6547},
pages = {1192-1196},
year = {2021},
doi = {10.1126/science.abg8102},
URL = {https://www.science.org/doi/abs/10.1126/science.abg8102},
eprint = {}}

@article{Else2019Review,
   author = "Else, Dominic V. and Monroe, Christopher and Nayak, Chetan and Yao, Norman Y.",
   title = "Discrete Time Crystals", 
   journal= "Annual Review of Condensed Matter Physics",
   year = "2020",
   volume = "11",
   number = "Volume 11, 2020",
   pages = "467-499",
   doi = "https://doi.org/10.1146/annurev-conmatphys-031119-050658",
   url = "https://www.annualreviews.org/content/journals/10.1146/annurev-conmatphys-031119-050658",
   publisher = "Annual Reviews",
   issn = "1947-5462",
   type = "Journal Article",
  }

@article{Keyserlingk2016,
  title = {Absolute stability and spatiotemporal long-range order in {F}loquet systems},
  author = {von Keyserlingk, C. W. and Khemani, Vedika and Sondhi, S. L.},
  journal = {Phys. Rev. B},
  volume = {94},
  issue = {8},
  pages = {085112},
  numpages = {11},
  year = {2016},
  month = {Aug},
  publisher = {American Physical Society},
  doi = {10.1103/PhysRevB.94.085112},
  url = {https://link.aps.org/doi/10.1103/PhysRevB.94.085112}
}

@article{Yao2017,
  title = {Discrete Time Crystals: Rigidity, Criticality, and Realizations},
  author = {Yao, N. Y. and Potter, A. C. and Potirniche, I.-D. and Vishwanath, A.},
  journal = {Phys. Rev. Lett.},
  volume = {118},
  issue = {3},
  pages = {030401},
  numpages = {6},
  year = {2017},
  month = {Jan},
  publisher = {American Physical Society},
  doi = {10.1103/PhysRevLett.118.030401},
  url = {https://link.aps.org/doi/10.1103/PhysRevLett.118.030401}
}

@article{Hohenberg1967,
  title = {Existence of Long-Range Order in One and Two Dimensions},
  author = {Hohenberg, P. C.},
  journal = {Phys. Rev.},
  volume = {158},
  issue = {2},
  pages = {383--386},
  numpages = {0},
  year = {1967},
  month = {Jun},
  publisher = {American Physical Society},
  doi = {10.1103/PhysRev.158.383},
  url = {https://link.aps.org/doi/10.1103/PhysRev.158.383}
}

@article{Haegeman,
	Author = {Haegeman, Jutho and Cirac, J. Ignacio and Osborne, Tobias J. and Pi\ifmmode \check{z}\else \v{z}\fi{}orn, Iztok and Verschelde, Henri and Verstraete, Frank},
	Doi = {10.1103/PhysRevLett.107.070601},
	Issue = {7},
	Journal = {Phys. Rev. Lett.},
	Month = {Aug},
	Numpages = {5},
	Pages = {070601},
	Publisher = {American Physical Society},
	Title = {Time-Dependent Variational Principle for Quantum Lattices},
	Url = {https://link.aps.org/doi/10.1103/PhysRevLett.107.070601},
	Volume = {107},
	Year = {2011}}

@article{Deutsch1991,
  title = {Quantum statistical mechanics in a closed system},
  author = {Deutsch, J. M.},
  journal = {Phys. Rev. A},
  volume = {43},
  issue = {4},
  pages = {2046--2049},
  numpages = {0},
  year = {1991},
  month = {Feb},
  publisher = {American Physical Society},
  doi = {10.1103/PhysRevA.43.2046},
  url = {https://link.aps.org/doi/10.1103/PhysRevA.43.2046}
}

@article{Srednecki1994,
  title = {Chaos and quantum thermalization},
  author = {Srednicki, Mark},
  journal = {Phys. Rev. E},
  volume = {50},
  issue = {2},
  pages = {888--901},
  numpages = {0},
  year = {1994},
  month = {Aug},
  publisher = {American Physical Society},
  doi = {10.1103/PhysRevE.50.888},
  url = {https://link.aps.org/doi/10.1103/PhysRevE.50.888}
}

@Article{Tasaki2019,
author={Tasaki, Hal},
title={Long-Range Order, ``Tower'' of States, and Symmetry Breaking in Lattice Quantum Systems},
journal={Journal of Statistical Physics},
year={2019},
month={Feb},
day={01},
volume={174},
number={4},
pages={735-761},
issn={1572-9613},
doi={10.1007/s10955-018-2193-8},
url={https://doi.org/10.1007/s10955-018-2193-8}
}

@article{ITensor,
	title={{The ITensor Software Library for Tensor Network Calculations}},
	author={Matthew Fishman and Steven R. White and E. Miles Stoudenmire},
	journal={SciPost Phys. Codebases},
	pages={4},
	year={2022},
	publisher={SciPost},
	doi={10.21468/SciPostPhysCodeb.4},
	url={https://scipost.org/10.21468/SciPostPhysCodeb.4},
}

@article{Mpemba1969,
	author = {E B Mpemba and D G Osborne},
	doi = {10.1088/0031-9120/4/3/312},
	journal = {Physics Education},
	month = {may},
	number = {3},
	pages = {172},
	publisher = {},
	title = {Cool?},
	url = {https://dx.doi.org/10.1088/0031-9120/4/3/312},
	volume = {4},
	year = {1969}
}

@article{Ares2023,
  pages = {2036},
  title = {Entanglement asymmetry as a probe of symmetry breaking},
  volume = {14},
  ISSN = {2041-1723},
  url = {http://dx.doi.org/10.1038/s41467-023-37747-8},
  DOI = {10.1038/s41467-023-37747-8},
  number = {1},
  journal = {Nature Communications},
  publisher = {Springer Science and Business Media LLC},
  author = {Ares,  Filiberto and Murciano,  Sara and Calabrese,  Pasquale},
  year = {2023},
  month = apr 
}

@article{Joshi2024,
  title = {Observing the Quantum {M}pemba Effect in Quantum Simulations},
  author = {Joshi, Lata Kh. and Franke, Johannes and Rath, Aniket and Ares, Filiberto and Murciano, Sara and Kranzl, Florian and Blatt, Rainer and Zoller, Peter and Vermersch, Beno\^{\i}t and Calabrese, Pasquale and Roos, Christian F. and Joshi, Manoj K.},
  journal = {Phys. Rev. Lett.},
  volume = {133},
  issue = {1},
  pages = {010402},
  numpages = {7},
  year = {2024},
  month = {Jul},
  publisher = {American Physical Society},
  doi = {10.1103/PhysRevLett.133.010402},
  url = {https://link.aps.org/doi/10.1103/PhysRevLett.133.010402}
}

@article{Rylands2024_2,
  title = {Microscopic Origin of the Quantum {M}pemba Effect in Integrable Systems},
  author = {Rylands, Colin and Klobas, Katja and Ares, Filiberto and Calabrese, Pasquale and Murciano, Sara and Bertini, Bruno},
  journal = {Phys. Rev. Lett.},
  volume = {133},
  issue = {1},
  pages = {010401},
  numpages = {6},
  year = {2024},
  month = {Jul},
  publisher = {American Physical Society},
  doi = {10.1103/PhysRevLett.133.010401},
  url = {https://link.aps.org/doi/10.1103/PhysRevLett.133.010401}
}

@article{Liu2024_3,
  title = {Symmetry Restoration and Quantum {M}pemba Effect in Symmetric Random Circuits},
  author = {Liu, Shuo and Zhang, Hao-Kai and Yin, Shuai and Zhang, Shi-Xin},
  journal = {Phys. Rev. Lett.},
  volume = {133},
  issue = {14},
  pages = {140405},
  numpages = {7},
  year = {2024},
  month = {Oct},
  publisher = {American Physical Society},
  doi = {10.1103/PhysRevLett.133.140405},
  url = {https://link.aps.org/doi/10.1103/PhysRevLett.133.140405}
}

@article{Liu2024_2,
   title={Symmetry restoration and quantum {M}pemba effect in many-body localization systems},
   volume={70},
   ISSN={2095-9273},
   url={http://dx.doi.org/10.1016/j.scib.2025.10.017},
   DOI={10.1016/j.scib.2025.10.017},
   number={23},
   journal={Science Bulletin},
   publisher={Elsevier BV},
   author={Liu, Shuo and Zhang, Hao-Kai and Yin, Shuai and Zhang, Shi-Xin and Yao, Hong},
   year={2025},
   month=Dec, pages={3991–3996} }

@article{dAlessio2016,
	Author = {Luca D'Alessio and Yariv Kafri and Anatoli Polkovnikov and Marcos Rigol},
	Doi = {10.1080/00018732.2016.1198134},
	Journal = {Adv. Phys.},
	Number = {3},
	Pages = {239},
	Publisher = {Taylor & Francis},
	Title = {From quantum chaos and eigenstate thermalization to statistical mechanics and thermodynamics},
	Url = {https://doi.org/10.1080/00018732.2016.1198134},
	Volume = {65},
	Year = {2016}}

@Article{Ueda2020,
	author={Ueda, Masahito},
	title={Quantum equilibration, thermalization and prethermalization in ultracold atoms},
	journal={Nature Reviews Physics},
	year={2020},
	month={Dec},
	day={01},
	volume={2},
	number={12},
	pages={669-681},
	issn={2522-5820},
	doi={10.1038/s42254-020-0237-x},
	url={https://doi.org/10.1038/s42254-020-0237-x}
}

@article{RigolNature,
	Author = {Rigol, Marcos and Dunjko, Vanja and Olshanii, Maxim},
	Date = {2008/04/17/print},
	Day = {17},
	Doi = {10.1038/nature06838},
	Isbn = {0028-0836},
	Journal = {Nature},
	L3 = {http://www.nature.com/nature/journal/v452/n7189/suppinfo/nature06838_S1.html},
	Month = {04},
	Number = {7189},
	Pages = {854--858},
	Publisher = {Nature Publishing Group},
	Title = {Thermalization and its mechanism for generic isolated quantum systems},
	Ty = {JOUR},
	Volume = {452},
	Year = {2008}}

@article{Mermin1966,
  title = {Absence of Ferromagnetism or Antiferromagnetism in One- or Two-Dimensional Isotropic {H}eisenberg Models},
  author = {Mermin, N. D. and Wagner, H.},
  journal = {Phys. Rev. Lett.},
  volume = {17},
  issue = {22},
  pages = {1133--1136},
  numpages = {0},
  year = {1966},
  month = {Nov},
  publisher = {American Physical Society},
  doi = {10.1103/PhysRevLett.17.1133},
  url = {https://link.aps.org/doi/10.1103/PhysRevLett.17.1133}
}

@article{Maghrebi2017,
  title = {Continuous Symmetry Breaking in {1D} Long-Range Interacting Quantum Systems},
  author = {Maghrebi, Mohammad F. and Gong, Zhe-Xuan and Gorshkov, Alexey V.},
  journal = {Phys. Rev. Lett.},
  volume = {119},
  issue = {2},
  pages = {023001},
  numpages = {6},
  year = {2017},
  month = {Jul},
  publisher = {American Physical Society},
  doi = {10.1103/PhysRevLett.119.023001},
  url = {https://link.aps.org/doi/10.1103/PhysRevLett.119.023001}
}

@article{Ren2020,
  title = {Entanglement and correlations in a one-dimensional quantum spin-$\frac{1}{2}$ chain with anisotropic power-law long-range interactions},
  author = {Ren, Jie and You, Wen-Long and Wang, Xiaoqun},
  journal = {Phys. Rev. B},
  volume = {101},
  issue = {9},
  pages = {094410},
  numpages = {8},
  year = {2020},
  month = {Mar},
  publisher = {American Physical Society},
  doi = {10.1103/PhysRevB.101.094410},
  url = {https://link.aps.org/doi/10.1103/PhysRevB.101.094410}
}

@book{sutherland2004beautiful,
	Author = {Sutherland, Bill},
	Publisher = {World Scientific Publishing Company},
	Title = {Beautiful models: 70 years of exactly solved quantum many-body problems},
    Address = {River Edge, NJ},
	Year = {2004}}

@article{Machado2020,
  title = {Long-Range Prethermal Phases of Nonequilibrium Matter},
  author = {Machado, Francisco and Else, Dominic V. and Kahanamoku-Meyer, Gregory D. and Nayak, Chetan and Yao, Norman Y.},
  journal = {Phys. Rev. X},
  volume = {10},
  issue = {1},
  pages = {011043},
  numpages = {34},
  year = {2020},
  month = {Feb},
  publisher = {American Physical Society},
  doi = {10.1103/PhysRevX.10.011043},
  url = {https://link.aps.org/doi/10.1103/PhysRevX.10.011043}
}

@article{Lerose2020,
  title = {Origin of the slow growth of entanglement entropy in long-range interacting spin systems},
  author = {Lerose, Alessio and Pappalardi, Silvia},
  journal = {Phys. Rev. Res.},
  volume = {2},
  issue = {1},
  pages = {012041},
  numpages = {7},
  year = {2020},
  month = {Feb},
  publisher = {American Physical Society},
  doi = {10.1103/PhysRevResearch.2.012041},
  url = {https://link.aps.org/doi/10.1103/PhysRevResearch.2.012041}
}

@article{LerosePappalardi,
  title = {Bridging entanglement dynamics and chaos in semiclassical systems},
  author = {Lerose, Alessio and Pappalardi, Silvia},
  journal = {Phys. Rev. A},
  volume = {102},
  issue = {3},
  pages = {032404},
  numpages = {25},
  year = {2020},
  month = {Sep},
  publisher = {American Physical Society},
  doi = {10.1103/PhysRevA.102.032404},
  url = {https://link.aps.org/doi/10.1103/PhysRevA.102.032404}
}

@article{Sciolla2013,
  title = {Quantum quenches, dynamical transitions, and off-equilibrium quantum criticality},
  author = {Sciolla, Bruno and Biroli, Giulio},
  journal = {Phys. Rev. B},
  volume = {88},
  issue = {20},
  pages = {201110},
  numpages = {5},
  year = {2013},
  month = {Nov},
  publisher = {American Physical Society},
  doi = {10.1103/PhysRevB.88.201110},
  url = {https://link.aps.org/doi/10.1103/PhysRevB.88.201110}
}

@article{Sciolla2010,
  title = {Quantum Quenches and Off-Equilibrium Dynamical Transition in the Infinite-Dimensional {Bose}-{Hubbard} Model},
  author = {Sciolla, Bruno and Biroli, Giulio},
  journal = {Phys. Rev. Lett.},
  volume = {105},
  issue = {22},
  pages = {220401},
  numpages = {4},
  year = {2010},
  month = {Nov},
  publisher = {American Physical Society},
  doi = {10.1103/PhysRevLett.105.220401},
  url = {https://link.aps.org/doi/10.1103/PhysRevLett.105.220401}
}

@article{Schafer2020,
   title={Tools for quantum simulation with ultracold atoms in optical lattices},
   volume={2},
   ISSN={2522-5820},
   url={http://dx.doi.org/10.1038/s42254-020-0195-3},
   DOI={10.1038/s42254-020-0195-3},
   number={8},
   journal={Nature Reviews Physics},
   publisher={Springer Science and Business Media LLC},
   author={Schäfer, Florian and Fukuhara, Takeshi and Sugawa, Seiji and Takasu, Yosuke and Takahashi, Yoshiro},
   year={2020},
   month=jul, pages={411–425} }

@article{LMG1,
title = {Validity of many-body approximation methods for a solvable model: (I). Exact solutions and perturbation theory},
journal = {Nuclear Physics},
volume = {62},
number = {2},
pages = {188-198},
year = {1965},
issn = {0029-5582},
doi = {https://doi.org/10.1016/0029-5582(65)90862-X},
url = {https://www.sciencedirect.com/science/article/pii/002955826590862X},
author = {H.J. Lipkin and N. Meshkov and A.J. Glick}
}

@article{LMG2,
title = {Validity of many-body approximation methods for a solvable model: (II). Linearization procedures},
journal = {Nuclear Physics},
volume = {62},
number = {2},
pages = {199-210},
year = {1965},
issn = {0029-5582},
doi = {https://doi.org/10.1016/0029-5582(65)90863-1},
url = {https://www.sciencedirect.com/science/article/pii/0029558265908631},
author = {N. Meshkov and A.J. Glick and H.J. Lipkin}
}

@article{Gross2017,
author = {Christian Gross  and Immanuel Bloch },
title = {Quantum simulations with ultracold atoms in optical lattices},
journal = {Science},
volume = {357},
number = {6355},
pages = {995-1001},
year = {2017},
doi = {10.1126/science.aal3837},
URL = {https://www.science.org/doi/abs/10.1126/science.aal3837}}

@article{Choi2020,
  title = {Robust Dynamic {H}amiltonian Engineering of Many-Body Spin Systems},
  author = {Choi, Joonhee and Zhou, Hengyun and Knowles, Helena S. and Landig, Renate and Choi, Soonwon and Lukin, Mikhail D.},
  journal = {Phys. Rev. X},
  volume = {10},
  issue = {3},
  pages = {031002},
  numpages = {27},
  year = {2020},
  month = {Jul},
  publisher = {American Physical Society},
  doi = {10.1103/PhysRevX.10.031002},
  url = {https://link.aps.org/doi/10.1103/PhysRevX.10.031002}
}

@article{Zhou2020,
  title = {Quantum Metrology with Strongly Interacting Spin Systems},
  author = {Zhou, Hengyun and Choi, Joonhee and Choi, Soonwon and Landig, Renate and Douglas, Alexander M. and Isoya, Junichi and Jelezko, Fedor and Onoda, Shinobu and Sumiya, Hitoshi and Cappellaro, Paola and Knowles, Helena S. and Park, Hongkun and Lukin, Mikhail D.},
  journal = {Phys. Rev. X},
  volume = {10},
  issue = {3},
  pages = {031003},
  numpages = {9},
  year = {2020},
  month = {Jul},
  publisher = {American Physical Society},
  doi = {10.1103/PhysRevX.10.031003},
  url = {https://link.aps.org/doi/10.1103/PhysRevX.10.031003}
}

@article{Haegeman2016,
  title = {Unifying time evolution and optimization with matrix product states},
  author = {Haegeman, Jutho and Lubich, Christian and Oseledets, Ivan and Vandereycken, Bart and Verstraete, Frank},
  journal = {Phys. Rev. B},
  volume = {94},
  issue = {16},
  pages = {165116},
  numpages = {10},
  year = {2016},
  month = {Oct},
  publisher = {American Physical Society},
  doi = {10.1103/PhysRevB.94.165116},
  url = {https://link.aps.org/doi/10.1103/PhysRevB.94.165116}
}

@article{Vanderstraeten2019,
  pages = {7},
   title={Tangent-space methods for uniform matrix product states},
   ISSN={2590-1990},
   url={http://dx.doi.org/10.21468/SciPostPhysLectNotes.7},
   DOI={10.21468/scipostphyslectnotes.7},
   journal={SciPost Physics Lecture Notes},
   publisher={Stichting SciPost},
   author={Vanderstraeten, Laurens and Haegeman, Jutho and Verstraete, Frank},
   year={2019},
   month=jan }

@article{Anderson1952,
  title = {An Approximate Quantum Theory of the Antiferromagnetic Ground State},
  author = {Anderson, P. W.},
  journal = {Phys. Rev.},
  volume = {86},
  issue = {5},
  pages = {694--701},
  numpages = {0},
  year = {1952},
  month = {Jun},
  publisher = {American Physical Society},
  doi = {10.1103/PhysRev.86.694},
  url = {https://link.aps.org/doi/10.1103/PhysRev.86.694}
}

@Article{ares2025,
author={Ares, Filiberto
and Calabrese, Pasquale
and Murciano, Sara},
title={The quantum {Mpemba} effects},
journal={Nature Reviews Physics},
year={2025},
month={Aug},
day={01},
volume={7},
number={8},
pages={451-460},
issn={2522-5820},
doi={10.1038/s42254-025-00838-0},
url={https://doi.org/10.1038/s42254-025-00838-0}
}

@article{Defenu2023,
  title = {Long-range interacting quantum systems},
  author = {Defenu, Nicol\`o and Donner, Tobias and Macr\`{\i}, Tommaso and Pagano, Guido and Ruffo, Stefano and Trombettoni, Andrea},
  journal = {Rev. Mod. Phys.},
  volume = {95},
  issue = {3},
  pages = {035002},
  numpages = {70},
  year = {2023},
  month = {Aug},
  publisher = {American Physical Society},
  doi = {10.1103/RevModPhys.95.035002},
  url = {https://link.aps.org/doi/10.1103/RevModPhys.95.035002}
}

@article{Chen_2023,
   title={Continuous symmetry breaking in a two-dimensional {R}ydberg array},
   volume={616},
   ISSN={1476-4687},
   url={http://dx.doi.org/10.1038/s41586-023-05859-2},
   DOI={10.1038/s41586-023-05859-2},
   number={7958},
   journal={Nature},
   publisher={Springer Science and Business Media LLC},
   author={Chen, Cheng and Bornet, Guillaume and Bintz, Marcus and Emperauger, Gabriel and Leclerc, Lucas and Liu, Vincent S. and Scholl, Pascal and Barredo, Daniel and Hauschild, Johannes and Chatterjee, Shubhayu and Schuler, Michael and Läuchli, Andreas M. and Zaletel, Michael P. and Lahaye, Thierry and Yao, Norman Y. and Browaeys, Antoine},
   year={2023},
   month=feb, pages={691–695} }

@article{Feng_2023,
   title={Continuous symmetry breaking in a trapped-ion spin chain},
   volume={623},
   ISSN={1476-4687},
   url={http://dx.doi.org/10.1038/s41586-023-06656-7},
   DOI={10.1038/s41586-023-06656-7},
   number={7988},
   journal={Nature},
   publisher={Springer Science and Business Media LLC},
   author={Feng, Lei and Katz, Or and Haack, Casey and Maghrebi, Mohammad and Gorshkov, Alexey V. and Gong, Zhexuan and Cetina, Marko and Monroe, Christopher},
   year={2023},
   month=nov, pages={713–717} }

@article{turkeshi2024,
  title = {Quantum {M}pemba Effect in Random Circuits},
  author = {Turkeshi, Xhek and Calabrese, Pasquale and De Luca, Andrea},
  journal = {Phys. Rev. Lett.},
  volume = {135},
  issue = {4},
  pages = {040403},
  numpages = {10},
  year = {2025},
  month = {Jul},
  publisher = {American Physical Society},
  doi = {10.1103/5d6p-8d1b},
  url = {https://link.aps.org/doi/10.1103/5d6p-8d1b}
}

@article{Yamashika2024,
  title = {Entanglement asymmetry and quantum {M}pemba effect in two-dimensional free-fermion systems},
  author = {Yamashika, Shion and Ares, Filiberto and Calabrese, Pasquale},
  journal = {Phys. Rev. B},
  volume = {110},
  issue = {8},
  pages = {085126},
  numpages = {16},
  year = {2024},
  month = {Aug},
  publisher = {American Physical Society},
  doi = {10.1103/PhysRevB.110.085126},
  url = {https://link.aps.org/doi/10.1103/PhysRevB.110.085126}
}

@article{Gong2016TopologicalPhases,
  title = {Topological phases with long-range interactions},
  author = {Gong, Z.-X. and Maghrebi, M. F. and Hu, A. and Wall, M. L. and Foss-Feig, M. and Gorshkov, A. V.},
  journal = {Phys. Rev. B},
  volume = {93},
  issue = {4},
  pages = {041102},
  numpages = {6},
  year = {2016},
  month = {Jan},
  publisher = {American Physical Society},
  doi = {10.1103/PhysRevB.93.041102},
  url = {https://link.aps.org/doi/10.1103/PhysRevB.93.041102}
}

@article{Gong2016Kaleidoscope,
  title = {Kaleidoscope of quantum phases in a long-range interacting spin-1 chain},
  author = {Gong, Z.-X. and Maghrebi, M. F. and Hu, A. and Foss-Feig, M. and Richerme, P. and Monroe, C. and Gorshkov, A. V.},
  journal = {Phys. Rev. B},
  volume = {93},
  issue = {20},
  pages = {205115},
  numpages = {12},
  year = {2016},
  month = {May},
  publisher = {American Physical Society},
  doi = {10.1103/PhysRevB.93.205115},
  url = {https://link.aps.org/doi/10.1103/PhysRevB.93.205115}
}

@article{Calabrese2009,
   title={Entanglement entropy and conformal field theory},
   volume={42},
   ISSN={1751-8121},
   url={http://dx.doi.org/10.1088/1751-8113/42/50/504005},
   DOI={10.1088/1751-8113/42/50/504005},
   number={50},
   journal={Journal of Physics A: Mathematical and Theoretical},
   publisher={IOP Publishing},
   author={Calabrese, Pasquale and Cardy, John},
   year={2009},
   month=dec, pages={504005} }

@article{Calabrese2004,
   title={Entanglement entropy and quantum field theory},
   volume={2004},
   ISSN={1742-5468},
   url={http://dx.doi.org/10.1088/1742-5468/2004/06/P06002},
   DOI={10.1088/1742-5468/2004/06/p06002},
   number={06},
   journal={Journal of Statistical Mechanics: Theory and Experiment},
   publisher={IOP Publishing},
   author={Pasquale Calabrese and John Cardy},
   year={2004},
   month=jun, pages={P06002} }

@article{Vodola2014,
  title = {Kitaev Chains with Long-Range Pairing},
  author = {Vodola, Davide and Lepori, Luca and Ercolessi, Elisa and Gorshkov, Alexey V. and Pupillo, Guido},
  journal = {Phys. Rev. Lett.},
  volume = {113},
  issue = {15},
  pages = {156402},
  numpages = {5},
  year = {2014},
  month = {Oct},
  publisher = {American Physical Society},
  doi = {10.1103/PhysRevLett.113.156402},
  url = {https://link.aps.org/doi/10.1103/PhysRevLett.113.156402}
}

@article{Vodola2015,
   title={Long-range {I}sing and {K}itaev models: phases, correlations and edge modes},
   volume={18},
   ISSN={1367-2630},
   url={http://dx.doi.org/10.1088/1367-2630/18/1/015001},
   DOI={10.1088/1367-2630/18/1/015001},
   number={1},
   journal={New Journal of Physics},
   publisher={IOP Publishing},
   author={Vodola, Davide and Lepori, Luca and Ercolessi, Elisa and Pupillo, Guido},
   year={2015},
   month=dec, pages={015001} }

@article{Ejima2015,
  title = {Comparative density-matrix renormalization group study of symmetry-protected topological phases in spin-1 chain and {B}ose-{H}ubbard models},
  author = {Ejima, Satoshi and Fehske, Holger},
  journal = {Phys. Rev. B},
  volume = {91},
  issue = {4},
  pages = {045121},
  numpages = {9},
  year = {2015},
  month = {Jan},
  publisher = {American Physical Society},
  doi = {10.1103/PhysRevB.91.045121},
  url = {https://link.aps.org/doi/10.1103/PhysRevB.91.045121}
}

@article{Alcaraz1992,
  title = {Critical behavior of anisotropic spin-S {H}eisenberg chains},
  author = {Alcaraz, F. C. and Moreo, A.},
  journal = {Phys. Rev. B},
  volume = {46},
  issue = {5},
  pages = {2896--2907},
  numpages = {0},
  year = {1992},
  month = {Aug},
  publisher = {American Physical Society},
  doi = {10.1103/PhysRevB.46.2896},
  url = {https://link.aps.org/doi/10.1103/PhysRevB.46.2896}
}

@article{Carollo2021,
  title = {Exponentially Accelerated Approach to Stationarity in {M}arkovian Open Quantum Systems through the {M}pemba Effect},
  author = {Carollo, Federico and Lasanta, Antonio and Lesanovsky, Igor},
  journal = {Phys. Rev. Lett.},
  volume = {127},
  issue = {6},
  pages = {060401},
  numpages = {6},
  year = {2021},
  month = {Aug},
  publisher = {American Physical Society},
  doi = {10.1103/PhysRevLett.127.060401},
  url = {https://link.aps.org/doi/10.1103/PhysRevLett.127.060401}
}

@Article{Zhang2025,
author={Zhang, Jie
and Xia, Gang
and Wu, Chun-Wang
and Chen, Ting
and Zhang, Qian
and Xie, Yi
and Su, Wen-Bo
and Wu, Wei
and Qiu, Cheng-Wei
and Chen, Ping-Xing
and Li, Weibin
and Jing, Hui
and Zhou, Yan-Li},
title={Observation of quantum strong {Mpemba} effect},
journal={Nat. Commun.},
year={2025},
month={Jan},
day={06},
volume={16},
number={1},
pages={301},
issn={2041-1723},
doi={10.1038/s41467-024-54303-0},
url={https://doi.org/10.1038/s41467-024-54303-0}
}

@article{Kochsiek2022,
  title = {Accelerating the approach of dissipative quantum spin systems towards stationarity through global spin rotations},
  author = {Kochsiek, Simon and Carollo, Federico and Lesanovsky, Igor},
  journal = {Phys. Rev. A},
  volume = {106},
  issue = {1},
  pages = {012207},
  numpages = {7},
  year = {2022},
  month = {Jul},
  publisher = {American Physical Society},
  doi = {10.1103/PhysRevA.106.012207},
  url = {https://link.aps.org/doi/10.1103/PhysRevA.106.012207}
}

@article{bao2022,
  pages = {150403},
   title={Accelerating Quantum Relaxation via Temporary Reset: A {M}pemba-Inspired Approach},
   volume={135},
   ISSN={1079-7114},
   url={http://dx.doi.org/10.1103/g94p-7421},
   DOI={10.1103/g94p-7421},
   number={15},
   journal={Physical Review Letters},
   publisher={American Physical Society (APS)},
   author={Bao, Ruicheng and Hou, Zhonghuai},
   year={2025},
   month=Oct }

@article{Ivander2023,
  title = {Hyperacceleration of quantum thermalization dynamics by bypassing long-lived coherences: An analytical treatment},
  author = {Ivander, Felix and Anto-Sztrikacs, Nicholas and Segal, Dvira},
  journal = {Phys. Rev. E},
  volume = {108},
  issue = {1},
  pages = {014130},
  numpages = {15},
  year = {2023},
  month = {Jul},
  publisher = {American Physical Society},
  doi = {10.1103/PhysRevE.108.014130},
  url = {https://link.aps.org/doi/10.1103/PhysRevE.108.014130}
}

@article{Zhou2023,
  title = {Accelerating relaxation through {Liouvillian} exceptional point},
  author = {Zhou, Yan-Li and Yu, Xiao-Die and Wu, Chun-Wang and Li, Xie-Qian and Zhang, Jie and Li, Weibin and Chen, Ping-Xing},
  journal = {Phys. Rev. Res.},
  volume = {5},
  issue = {4},
  pages = {043036},
  numpages = {13},
  year = {2023},
  month = {Oct},
  publisher = {American Physical Society},
  doi = {10.1103/PhysRevResearch.5.043036},
  url = {https://link.aps.org/doi/10.1103/PhysRevResearch.5.043036}
}

@article{Chatterjee2023,
  title = {Quantum {Mpemba} Effect in a Quantum Dot with Reservoirs},
  author = {Chatterjee, Amit Kumar and Takada, Satoshi and Hayakawa, Hisao},
  journal = {Phys. Rev. Lett.},
  volume = {131},
  issue = {8},
  pages = {080402},
  numpages = {6},
  year = {2023},
  month = {Aug},
  publisher = {American Physical Society},
  doi = {10.1103/PhysRevLett.131.080402},
  url = {https://link.aps.org/doi/10.1103/PhysRevLett.131.080402}
}

@article{Wang2024,
  title = {Mpemba effects in nonequilibrium open quantum systems},
  author = {Wang, Xuanhua and Wang, Jin},
  journal = {Phys. Rev. Res.},
  volume = {6},
  issue = {3},
  pages = {033330},
  numpages = {7},
  year = {2024},
  month = {Sep},
  publisher = {American Physical Society},
  doi = {10.1103/PhysRevResearch.6.033330},
  url = {https://link.aps.org/doi/10.1103/PhysRevResearch.6.033330}
}

@article{Liu2024_4,
  title = {Speeding up quantum heat engines by the {Mpemba} effect},
  author = {Liu, Dehua and Yuan, Jiehong and Ruan, Huilin and Xu, Yang and Luo, Shaolin and He, Jizhou and He, Xian and Ma, Yongli and Wang, Jianhui},
  journal = {Phys. Rev. A},
  volume = {110},
  issue = {4},
  pages = {042218},
  numpages = {12},
  year = {2024},
  month = {Oct},
  publisher = {American Physical Society},
  doi = {10.1103/PhysRevA.110.042218},
  url = {https://link.aps.org/doi/10.1103/PhysRevA.110.042218}
}

@article{Longhi2024,
author = {Stefano Longhi},
journal = {Opt. Lett.},
number = {18},
pages = {5188--5191},
publisher = {Optica Publishing Group},
title = {Photonic {Mpemba} effect},
volume = {49},
month = {Sep},
year = {2024},
url = {https://opg.optica.org/ol/abstract.cfm?URI=ol-49-18-5188},
doi = {10.1364/OL.532503},
}

@article{wang2024mpembameetsquantumchaos,
  title = {Thermalization and {M}pemba-like patterns in effective temperature dynamics of strongly coupled dissipative quantum chaotic systems},
  author = {Wang, Xuanhua and Su, Jie and Wang, Jin},
  journal = {Phys. Rev. B},
  volume = {113},
  issue = {4},
  pages = {045119},
  numpages = {11},
  year = {2026},
  month = {Jan},
  publisher = {American Physical Society},
  doi = {10.1103/gjc2-8fkb},
  url = {https://link.aps.org/doi/10.1103/gjc2-8fkb}
}

@article{furtado2024,
title = {Enhanced quantum Mpemba effect with squeezed thermal reservoirs},
journal = {Annals of Physics},
volume = {480},
pages = {170135},
year = {2025},
issn = {0003-4916},
doi = {https://doi.org/10.1016/j.aop.2025.170135},
url = {https://www.sciencedirect.com/science/article/pii/S0003491625002179},
author = {J. Furtado and Alan C. Santos}
}

@article{qian2024,
  title = {Intrinsic quantum Mpemba effect in Markovian systems and quantum circuits},
  author = {Qian, Dongheng and Wang, Huan and Wang, Jing},
  journal = {Phys. Rev. B},
  volume = {111},
  issue = {22},
  pages = {L220304},
  numpages = {6},
  year = {2025},
  month = {Jun},
  publisher = {American Physical Society},
  doi = {10.1103/qj8n-k5j2},
  url = {https://link.aps.org/doi/10.1103/qj8n-k5j2}
}

@article{Bettmann2025,
  title = {Information geometry approach to quantum stochastic thermodynamics},
  author = {Bettmann, Laetitia P. and Goold, John},
  journal = {Phys. Rev. E},
  volume = {111},
  issue = {1},
  pages = {014133},
  numpages = {11},
  year = {2025},
  month = {Jan},
  publisher = {American Physical Society},
  doi = {10.1103/PhysRevE.111.014133},
  url = {https://link.aps.org/doi/10.1103/PhysRevE.111.014133}
}

@article{Dong2025,
  title = {Quantum {Mpemba} effect of localization in the dissipative mosaic model},
  author = {Dong, J. W. and Mu, H. F. and Qin, M. and Cui, H. T.},
  journal = {Phys. Rev. A},
  volume = {111},
  issue = {2},
  pages = {022215},
  numpages = {9},
  year = {2025},
  month = {Feb},
  publisher = {American Physical Society},
  doi = {10.1103/PhysRevA.111.022215},
  url = {https://link.aps.org/doi/10.1103/PhysRevA.111.022215}
}

@article{Nava2024,
  title = {Mpemba Effects in Open Nonequilibrium Quantum Systems},
  author = {Nava, Andrea and Egger, Reinhold},
  journal = {Phys. Rev. Lett.},
  volume = {133},
  issue = {13},
  pages = {136302},
  numpages = {7},
  year = {2024},
  month = {Sep},
  publisher = {American Physical Society},
  doi = {10.1103/PhysRevLett.133.136302},
  url = {https://link.aps.org/doi/10.1103/PhysRevLett.133.136302}
}

@article{medina2024,
  title = {Anomalous Discharging of Quantum Batteries: The Ergotropic {Mpemba} Effect},
  author = {Medina, Ivan and Culhane, Ois\'{\i}n and Binder, Felix C. and Landi, Gabriel T. and Goold, John},
  journal = {Phys. Rev. Lett.},
  volume = {134},
  issue = {22},
  pages = {220402},
  numpages = {6},
  year = {2025},
  month = {Jun},
  publisher = {American Physical Society},
  doi = {10.1103/PhysRevLett.134.220402},
  url = {https://link.aps.org/doi/10.1103/PhysRevLett.134.220402}
}

@article{kheirandish2024,
title = {The {Mpemba} effect in quantum oscillating and two-level systems},
journal = {Physics Letters A},
volume = {559},
pages = {130915},
year = {2025},
issn = {0375-9601},
doi = {https://doi.org/10.1016/j.physleta.2025.130915},
url = {https://www.sciencedirect.com/science/article/pii/S0375960125006954},
author = {F. Kheirandish and N. Cheraghpour and A. Moradian}
}

@article{graf2025,
doi = {10.1088/1361-648X/adc3e3},
url = {https://doi.org/10.1088/1361-648X/adc3e3},
year = {2025},
month = {apr},
publisher = {IOP Publishing},
volume = {37},
number = {19},
pages = {195302},
author = {Graf, Juliane and Splettstoesser, Janine and Monsel, Juliette},
title = {Role of electron–electron interaction in the Mpemba effect in quantum dots},
journal = {Journal of Physics: Condensed Matter}
}

@article{Zatsaryna2025,
  title = {Green's function approach to {J}osephson dot dynamics and application to quantum {M}pemba effects},
  author = {Zatsarynna, Kateryna and Nava, Andrea and Egger, Reinhold and Zazunov, Alex},
  journal = {Phys. Rev. B},
  volume = {111},
  issue = {10},
  pages = {104506},
  numpages = {19},
  year = {2025},
  month = {Mar},
  publisher = {American Physical Society},
  doi = {10.1103/PhysRevB.111.104506},
  url = {https://link.aps.org/doi/10.1103/PhysRevB.111.104506}
}

@article{Strachan2024,
  pages = {220403},
   title={Non-{M}arkovian Quantum {M}pemba Effect},
   volume={134},
   ISSN={1079-7114},
   url={http://dx.doi.org/10.1103/PhysRevLett.134.220403},
   DOI={10.1103/physrevlett.134.220403},
   number={22},
   journal={Physical Review Letters},
   publisher={American Physical Society (APS)},
   author={Strachan, David J. and Purkayastha, Archak and Clark, Stephen R.},
   year={2025},
   month=jun }

@article{wang2024goingquantummarkovianityreality,
      title={Going beyond quantum {M}arkovianity and back to reality: An exact master equation study}, 
      author={Zhao-Ming Wang and S. L. Wu and Mark S. Byrd and Lian-Ao Wu},
      year={2024},
      journal={arXiv:2411.17197},
      url={https://arxiv.org/abs/2411.17197}, 
}

@article{Murciano2024,
doi = {10.1088/1742-5468/ad17b4},
url = {https://dx.doi.org/10.1088/1742-5468/ad17b4},
year = {2024},
month = {jan},
publisher = {IOP Publishing},
volume = {2024},
number = {1},
pages = {013103},
author = {Murciano, Sara and Ares, Filiberto and Klich, Israel and Calabrese, Pasquale},
title = {Entanglement asymmetry and quantum {M}pemba effect in the {XY} spin chain},
journal = {J. Stat. Mech.: Theory Exp.}
}

@article{Rylands2024,
doi = {10.1088/1742-5468/ad97b3},
url = {https://dx.doi.org/10.1088/1742-5468/ad97b3},
year = {2024},
month = {dec},
publisher = {IOP Publishing},
volume = {2024},
number = {12},
pages = {123102},
author = {Rylands, Colin and Vernier, Eric and Calabrese, Pasquale},
title = {Dynamical symmetry restoration in the {H}eisenberg spin chain},
journal = {J. Stat. Mech.: Theory Exp.}
}

@article{yamashika2024quenchingsuperfluidfreebosons,
  title = {Quenching from superfluid to free bosons in two dimensions: Entanglement, symmetries, and the quantum Mpemba effect},
  author = {Yamashika, Shion and Calabrese, Pasquale and Ares, Filiberto},
  journal = {Phys. Rev. A},
  volume = {111},
  issue = {4},
  pages = {043304},
  numpages = {28},
  year = {2025},
  month = {Apr},
  publisher = {American Physical Society},
  doi = {10.1103/PhysRevA.111.043304},
  url = {https://link.aps.org/doi/10.1103/PhysRevA.111.043304}
}

@article{klobas2024,
  title = {Translation symmetry restoration under random unitary dynamics},
  author = {Klobas, Katja and Rylands, Colin and Bertini, Bruno},
  journal = {Phys. Rev. B},
  volume = {111},
  issue = {14},
  pages = {L140304},
  numpages = {6},
  year = {2025},
  month = {Apr},
  publisher = {American Physical Society},
  doi = {10.1103/PhysRevB.111.L140304},
  url = {https://link.aps.org/doi/10.1103/PhysRevB.111.L140304}
}

@article{yu2025,
  doi = {10.1088/0256-307X/42/11/110602},
  number = {11},
  volume = {42},
  pages = {110602},
	author={Yu, Hui and Li, Zi-Xiang and Zhang, Shi-Xin},
	title={Symmetry Breaking Dynamics in Quantum Many-Body Systems},
	journal={Chinese Physics Letters},
	url={http://iopscience.iop.org/article/10.1088/0256-307X/42/11/110602},
	year={2025}
}

@article{ares2025entanglementasymmetrydynamicsrandom,
  title = {Entanglement asymmetry dynamics in random quantum circuits},
  author = {Ares, Filiberto and Murciano, Sara and Calabrese, Pasquale and Piroli, Lorenzo},
  journal = {Phys. Rev. Res.},
  volume = {7},
  issue = {3},
  pages = {033135},
  numpages = {21},
  year = {2025},
  month = {Aug},
  publisher = {American Physical Society},
  doi = {10.1103/m3np-p5xj},
  url = {https://link.aps.org/doi/10.1103/m3np-p5xj}
}

@article{Klobas2024Rule54,
doi = {10.1088/1751-8121/ad91fd},
url = {https://dx.doi.org/10.1088/1751-8121/ad91fd},
year = {2024},
month = {nov},
publisher = {IOP Publishing},
volume = {57},
number = {50},
pages = {505001},
author = {Klobas, Katja},
title = {Non-equilibrium dynamics of symmetry-resolved entanglement and entanglement asymmetry: exact asymptotics in {R}ule 54},
journal = {J. Phys. A: Math. Theor.}
}

@article{Foligno2025,
  title = {Nonequilibrium Dynamics of Charged Dual-Unitary Circuits},
  author = {Foligno, Alessandro and Calabrese, Pasquale and Bertini, Bruno},
  journal = {PRX Quantum},
  volume = {6},
  issue = {1},
  pages = {010324},
  numpages = {34},
  year = {2025},
  month = {Feb},
  publisher = {American Physical Society},
  doi = {10.1103/PRXQuantum.6.010324},
  url = {https://link.aps.org/doi/10.1103/PRXQuantum.6.010324}
}

@article{Bull2022,
  title = {Tuning between Continuous Time Crystals and Many-Body Scars in Long-Range {XYZ} Spin Chains},
  author = {Bull, Kieran and Hallam, Andrew and Papi\ifmmode \acute{c}\else \'{c}\fi{}, Zlatko and Martin, Ivar},
  journal = {Phys. Rev. Lett.},
  volume = {129},
  issue = {14},
  pages = {140602},
  numpages = {7},
  year = {2022},
  month = {Sep},
  publisher = {American Physical Society},
  doi = {10.1103/PhysRevLett.129.140602},
  url = {https://link.aps.org/doi/10.1103/PhysRevLett.129.140602}
}

@article{Else2017,
  title = {Prethermal Phases of Matter Protected by Time-Translation Symmetry},
  author = {Else, Dominic V. and Bauer, Bela and Nayak, Chetan},
  journal = {Phys. Rev. X},
  volume = {7},
  issue = {1},
  pages = {011026},
  numpages = {21},
  year = {2017},
  month = {Mar},
  publisher = {American Physical Society},
  doi = {10.1103/PhysRevX.7.011026},
  url = {https://link.aps.org/doi/10.1103/PhysRevX.7.011026}
}

@article{Kac1963,
       author = {{Kac}, M. and {Uhlenbeck}, G.~E. and {Hemmer}, P.~C.},
        title = {On the van der {W}aals Theory of the Vapor-Liquid Equilibrium. {I}. Discussion of a One-Dimensional Model},
      journal = {Journal of Mathematical Physics},
         year = 1963,
        month = feb,
       volume = {4},
       number = {2},
        pages = {216-228},
          doi = {10.1063/1.1703946},
       adsurl = {https://ui.adsabs.harvard.edu/abs/1963JMP.....4..216K}
}

@article{Abanin_2017,
   title={A Rigorous Theory of Many-Body Prethermalization for Periodically Driven and Closed Quantum Systems},
   volume={354},
   ISSN={1432-0916},
   url={http://dx.doi.org/10.1007/s00220-017-2930-x},
   DOI={10.1007/s00220-017-2930-x},
   number={3},
   journal={Communications in Mathematical Physics},
   publisher={Springer Science and Business Media LLC},
   author={Abanin, Dmitry and De Roeck, Wojciech and Ho, Wen Wei and Huveneers, François},
   year={2017},
   month=jun, pages={809–827} }

@article{Ares2025FreeFermionsMixed,
  title = {Quantum {M}pemba effect in free-fermionic mixed states},
  author = {Ares, Filiberto and Vitale, Vittorio and Murciano, Sara},
  journal = {Phys. Rev. B},
  volume = {111},
  issue = {10},
  pages = {104312},
  numpages = {17},
  year = {2025},
  month = {Mar},
  publisher = {American Physical Society},
  doi = {10.1103/PhysRevB.111.104312},
  url = {https://link.aps.org/doi/10.1103/PhysRevB.111.104312}
}

@article{Caceffo2024,
doi = {10.1088/1742-5468/ad4537},
url = {https://dx.doi.org/10.1088/1742-5468/ad4537},
year = {2024},
month = {jun},
publisher = {IOP Publishing},
volume = {2024},
number = {6},
pages = {063103},
author = {Caceffo, Fabio and Murciano, Sara and Alba, Vincenzo},
title = {Entangled multiplets, asymmetry, and quantum {M}pemba effect in dissipative systems},
journal = {J. Stat. Mech.: Theory Exp.}
}

@article{bhore2025,
  title = {Quantum {Mpemba} effect without global symmetries},
  author = {Bhore, Tanmay and Su, Lei and Martin, Ivar and Clerk, Aashish A. and Papi\ifmmode \acute{c}\else \'{c}\fi{}, Zlatko},
  journal = {Phys. Rev. B},
  volume = {112},
  issue = {12},
  pages = {L121109},
  numpages = {6},
  year = {2025},
  month = {Sep},
  publisher = {American Physical Society},
  doi = {10.1103/1td3-2vwf},
  url = {https://link.aps.org/doi/10.1103/1td3-2vwf}
}

@article{fishman2025itensor-r0.3,
	title={{Codebase release 0.3 for ITensor}},
	author={Matthew Fishman and Steven R. White and E. Miles Stoudenmire},
	journal={SciPost Phys. Codebases},
	pages={4-r0.3},
	year={2022},
	publisher={SciPost},
	doi={10.21468/SciPostPhysCodeb.4-r0.3},
	url={https://scipost.org/10.21468/SciPostPhysCodeb.4-r0.3}
}

@article{Schneider2022,
  title = {Entanglement spectrum and quantum phase diagram of the long-range XXZ chain},
  author = {Schneider, J. T. and Thomson, S. J. and Sanchez-Palencia, L.},
  journal = {Phys. Rev. B},
  volume = {106},
  issue = {1},
  pages = {014306},
  numpages = {14},
  year = {2022},
  month = {Jul},
  publisher = {American Physical Society},
  doi = {10.1103/PhysRevB.106.014306},
  url = {https://link.aps.org/doi/10.1103/PhysRevB.106.014306}
}

@article{Yamashika2025,
  pages = {090402},
   title={Quantum {M}pemba Effect in Long-Range Spin Systems},
   volume={136},
   ISSN={1079-7114},
   url={http://dx.doi.org/10.1103/52y5-8kl2},
   DOI={10.1103/52y5-8kl2},
   number={9},
   journal={Physical Review Letters},
   publisher={American Physical Society (APS)},
   author={Yamashika, Shion and Ares, Filiberto},
   year={2026},
   month=Mar }

@article{Zhao_2023,
  pages = {59},
   title={Finite-temperature critical behaviors in 2D long-range quantum {H}eisenberg model},
   volume={8},
   ISSN={2397-4648},
   url={http://dx.doi.org/10.1038/s41535-023-00591-6},
   DOI={10.1038/s41535-023-00591-6},
   number={1},
   journal={npj Quantum Materials},
   publisher={Springer Science and Business Media LLC},
   author={Zhao, Jiarui and Song, Menghan and Qi, Yang and Rong, Junchen and Meng, Zi Yang},
   year={2023},
   month=oct }

@article{Yu2025_2,
  pages = {17},
   title={Quantum {M}pemba effects from symmetry perspectives},
   volume={35},
   ISSN={2309-4710},
   url={http://dx.doi.org/10.1007/s43673-025-00157-7},
   DOI={10.1007/s43673-025-00157-7},
   number={1},
   journal={AAPPS Bulletin},
   publisher={Springer Science and Business Media LLC},
   author={Yu, Hui and Liu, Shuo and Zhang, Shi-Xin},
   year={2025},
   month=jul }

@article{ares2025simplerprobeqme,
   title={A simpler probe of the quantum {M}pemba effect in closed systems},
   volume={58},
   ISSN={1751-8121},
   url={http://dx.doi.org/10.1088/1751-8121/ae0d18},
   DOI={10.1088/1751-8121/ae0d18},
   number={44},
   journal={Journal of Physics A: Mathematical and Theoretical},
   publisher={IOP Publishing},
   author={Ares, Filiberto and Rylands, Colin and Calabrese, Pasquale},
   year={2025},
   month=Oct, pages={445302} }

@article{Mondal2025,
  pages = {022211},
   title={{M}pemba effect in self-contained quantum refrigerators: Accelerated cooling},
   volume={113},
   ISSN={2469-9934},
   url={http://dx.doi.org/10.1103/8fp4-15mq},
   DOI={10.1103/8fp4-15mq},
   number={2},
   journal={Physical Review A},
   publisher={American Physical Society (APS)},
   author={Mondal, Sayan and Sen, Ujjwal},
   year={2026},
   month=Feb }

@article{Summer2025,
  pages = {011065},
   title={Resource-Theoretical Unification of {M}pemba Effects: Classical and Quantum},
   volume={16},
   ISSN={2160-3308},
   url={http://dx.doi.org/10.1103/rbt4-psfd},
   DOI={10.1103/rbt4-psfd},
   number={1},
   journal={Physical Review X},
   publisher={American Physical Society (APS)},
   author={Summer, Alessandro and Moroder, Mattia and Bettmann, Laetitia P. and Turkeshi, Xhek and Marvian, Iman and Goold, John},
   year={2026},
   month=Mar }

@article{Sreejith2025,
  pages = {214322},
   title={Signatures of quantum chaos and complexity in the {I}sing model on random graphs},
   volume={113},
   ISSN={2469-9969},
   url={http://dx.doi.org/10.1103/vt39-tvrs},
   DOI={10.1103/vt39-tvrs},
   number={21},
   journal={Physical Review B},
   publisher={American Physical Society (APS)},
   author={Sreejith, G. J. and Manna, Sandipan},
   year={2026},
   month=jun }

@article{Xu2025,
  pages = {010402},
   title={Observation and Modulation of the Quantum {M}pemba Effect on a Superconducting Quantum Processor},
   volume={137},
   ISSN={1079-7114},
   url={http://dx.doi.org/10.1103/951q-j8kq},
   DOI={10.1103/951q-j8kq},
   number={1},
   journal={Physical Review Letters},
   publisher={American Physical Society (APS)},
   author={Xu, Yueshan and Fang, Cai-Ping and Chen, Bing-Jie and Wang, Ming-Chuan and Ge, Zi-Yong and Shi, Yun-Hao and Liu, Yu and Deng, Cheng-Lin and Zhao, Kui and Liu, Zheng-He and Li, Tian-Ming and Li, Hao and Wang, Ziting and Liang, Gui-Han and Feng, Da’er and Guo, Xue-Yi and Gu, Xu-Yang and He, Yang and Liu, Hao-Tian and Mei, Zheng-Yang and Xiao, Yongxi and Yan, Yu and Yu, Yi-Han and Yuan, Wei-Ping and Zhang, Jia-Chi and Wang, Zheng-An and Liu, Gangqin and Song, Xiaohui and Tian, Ye and Zhang, Yu-Ran and Zhang, Shi-Xin and Huang, Kaixuan and Xiang, Zhongcheng and Zheng, Dongning and Xu, Kai and Fan, Heng},
   year={2026},
   month=jun }

@article{ma2025,
  pages = {214310},
   title={Quantum {M}pemba effect in parity-time symmetric systems},
   volume={112},
   ISSN={2469-9969},
   url={http://dx.doi.org/10.1103/cpnh-myg9},
   DOI={10.1103/cpnh-myg9},
   number={21},
   journal={Physical Review B},
   publisher={American Physical Society (APS)},
   author={Ma, Wanchen and Liu, Junjie},
   year={2025},
   month=Dec }

@article{wei2025,
  author  = {Wei, Zijun and Xu, Mingdi and Jiang, Xiang-Ping and Hu, Haiping and Pan, Lei},
  title   = {Quantum {M}pemba effect in dissipative spin chains at criticality},
  journal = {Science China Physics, Mechanics \& Astronomy},
  year    = {2026},
  volume  = {69},
  number  = {4},
  pages   = {240315},
  doi     = {10.1007/s11433-025-2878-4},
  issn    = {1869-1927},
  url     = {https://doi.org/10.1007/s11433-025-2878-4}
}

@article{Parez2025,
   title={Reduced fidelities for free fermions out of equilibrium: from dynamical quantum phase transitions to {M}pemba effect},
   volume={2026},
   ISSN={1742-5468},
   url={http://dx.doi.org/10.1088/1742-5468/ae3380},
   DOI={10.1088/1742-5468/ae3380},
   number={1},
   journal={Journal of Statistical Mechanics: Theory and Experiment},
   publisher={IOP Publishing},
   author={Parez, Gilles and Alba, Vincenzo},
   year={2026},
   month=Jan, pages={013103} }

@article{Fujimura2025,
  author  = {Fujimura, Harunobu and Shimamori, Soichiro},
  title   = {Entanglement asymmetry and quantum {M}pemba effect for non-{A}belian global symmetry},
  journal = {Journal of High Energy Physics},
  year    = {2026},
  volume  = {2026},
  number  = {3},
  pages   = {244},
  doi     = {10.1007/JHEP03(2026)244},
  issn    = {1029-8479},
  url     = {https://doi.org/10.1007/JHEP03(2026)244}
}

@article{Ali2025,
  pages = {e70089},
   title={Quantum {M}pemba Effect in a Four‐Site Bose–Hubbard Model},
   volume={74},
   ISSN={1521-3978},
   url={http://dx.doi.org/10.1002/prop.70089},
   DOI={10.1002/prop.70089},
   number={3},
   journal={Fortschritte der Physik},
   publisher={Wiley},
   author={Ali, Asad and Zad, Hamid Arian and Hussain, Muhammad Irtiza and Al‐Kuwari, Saif and Kuniyil, Hashir and Rahim, Muhammad Talha and Jaščur, Michal and Haddadi, Saeed},
   year={2026},
   month=Mar }

@article{Li2025,
  title = {Ergotropic {M}pemba effect in non-{M}arkovian quantum systems},
  author = {Li, Yan and Li, Wenlin and Li, Xingli},
  journal = {Phys. Rev. A},
  volume = {112},
  issue = {3},
  pages = {032209},
  numpages = {15},
  year = {2025},
  month = {Sep},
  publisher = {American Physical Society},
  doi = {10.1103/5xrr-x2rm},
  url = {https://link.aps.org/doi/10.1103/5xrr-x2rm}
}

@misc{yamashika2025_2,
      title={Quantum {F}isher Information as a Measure of Symmetry Breaking in Quantum Many-Body Systems}, 
      author={Shion Yamashika and Shimpei Endo and Hiroyasu Tajima},
      year={2025},
      eprint={2509.07468},
      archivePrefix={arXiv},
      primaryClass={cond-mat.stat-mech},
      url={https://arxiv.org/abs/2509.07468}, 
}

@misc{chatterjee2025,
      title={Direct Experimental Observation of Quantum {M}pemba Effect without Bath Engineering}, 
      author={Arijit Chatterjee and Sakil Khan and Sachin Jain and T S Mahesh},
      year={2025},
      eprint={2509.13451},
      archivePrefix={arXiv},
      primaryClass={quant-ph},
      url={https://arxiv.org/abs/2509.13451}, 
}

@article{Pirvu2010,
   title={Matrix product operator representations},
   volume={12},
   ISSN={1367-2630},
   url={http://dx.doi.org/10.1088/1367-2630/12/2/025012},
   DOI={10.1088/1367-2630/12/2/025012},
   number={2},
   journal={New Journal of Physics},
   publisher={IOP Publishing},
   author={Pirvu, B and Murg, V and Cirac, J I and Verstraete, F},
   year={2010},
   month=Feb, pages={025012} }

@misc{dataset2026,
  author = {Hallam, Andrew and Yusuf, Matthew and Clerk, Aashish A. and Martin, Ivar and Papi\'c, Zlatko},
  title  = {10.5518/1825},
  publisher = {University of Leeds},
  doi    = {10.5518/1825},
  url    = {https://doi.org/10.5518/1825}
}

\end{document}


\title{Supplementary Material for ``Tunable quantum Mpemba effect in long-range interacting systems''
}

\author{Andrew Hallam \orcidlink{0000-0003-2288-7661}}
\thanks{These authors contributed equally.}
\affiliation{School of Physics and Astronomy, University of Leeds, Leeds LS2 9JT, United Kingdom}

\author{Matthew Yusuf \orcidlink{0009-0005-7080-8499}}
\thanks{These authors contributed equally.} 
\affiliation{School of Physics and Astronomy, University of Leeds, Leeds LS2 9JT, United Kingdom}

\author{Aashish A. Clerk}
\affiliation{Pritzker School of Molecular Engineering, University of Chicago, Chicago, Illinois 60637, USA}

\author{Ivar Martin}
\affiliation{Materials Science Division, Argonne National Laboratory, Lemont, Illinois 60439, USA}
\affiliation{Department of Physics, University of Chicago, Chicago, Illinois 60637, USA}

\author{Zlatko Papi\'c \orcidlink{0000-0002-8451-2235}}
\affiliation{School of Physics and Astronomy, University of Leeds, Leeds LS2 9JT, United Kingdom}

\date{November 10, 2026}
\begin{abstract}
In this Supplementary Material, we provide further details of trace distance as an alternative measure of the quantum Mpemba effect (QME), additional numerical results and finite-size scaling analysis, and the study of the SU(2)-symmetric point in the XYZ model.
\end{abstract}
\maketitle

\section{Trace distance as a probe of the QME}\label{app::trace}

\begin{figure}[tb]
    \centering
    \includegraphics[width=0.85\linewidth]{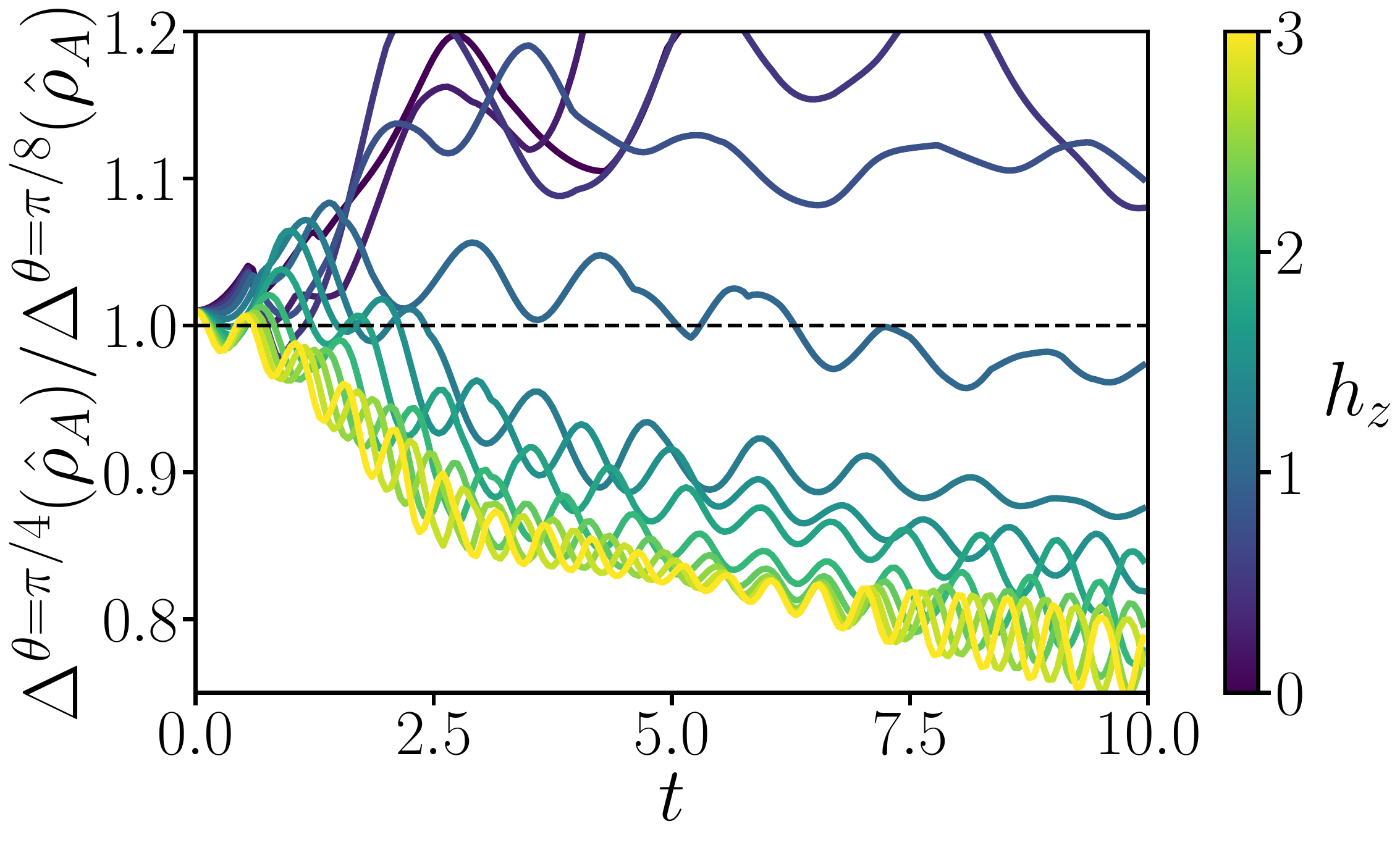}
    \caption{Ratio of trace distances for the initial states defined in Eq.~(9) of the main text at $\theta_1=\pi/4$ and $\theta_2=\pi/8$ evolved under the XYZ Hamiltonian [Eq.~(1) of the main text] with $J_x=-0.5$, $J_y=-1.5$, $J_z=-0.75$ and $\alpha=4$, with varying magnetic field strengths $h_z$ (color bar). Results are for a finite system of $N=100$ spins and subsystem of $N_A=4$ spins. The thermal density matrix was constructed using a smaller system of $14$ spins. The ratio crossing $1$ indicates the QME has occurred. Similar to Fig.~2(e) in the main text, no QME is observed at low fields, but increasing $h_z$ induces the QME. The dynamics data were obtained using TDVP for finite MPS, with the finite systems results using max bond dimension $\chi=100$. 
    }
    \label{fig:tracedist}
\end{figure}

\begin{figure}[bt]
    \centering
    \includegraphics[width=\linewidth]{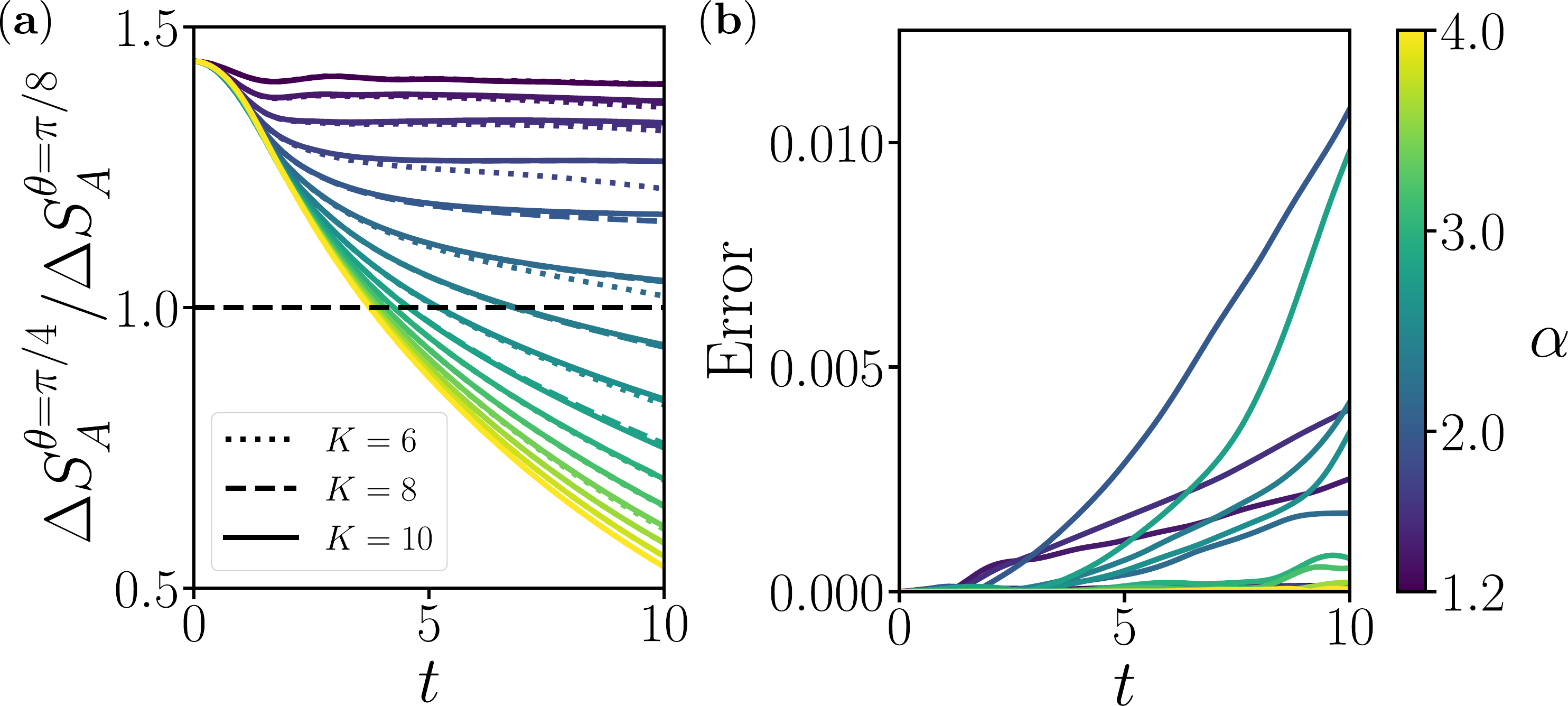}
    \caption{Convergence of the dynamics with the number of
    exponential terms $K$ used to approximate the power-law
    interaction in Eq.~\ref{eq:sumofexp}. (a) Ratio of the entanglement asymmetry
    $\Delta S_A^{\theta=\pi/4}/\Delta S_A^{\theta=\pi/8}$ for the
    initial states in Eq.~(9) of the main text, evolved under the
    XXZ Hamiltonian [Eq.~(3) of the main text] with $J_x = J_y =
    -1$, $J_z = -0.75$, for various values of $\alpha$ (color bar),
    computed with $K=6$ (dotted), $K=8$ (dashed), and $K=10$ (solid)
    exponential terms. (b) Relative error of the $K=8$ approximation
    with respect to $K=10$, Eq.~(\ref{eq:relerr}), 
    for the curves in panel (a). The error
    stays below $\sim 1\%$ at all times, confirming that value $K=8$ used throughout this work is sufficiently
    converged.}
    \label{fig:numExp}
\end{figure}

An alternative measure for the QME is the trace distance~\cite{ares2025} 
\begin{equation}\label{eq:tracedist}
    \Delta(\hat{\rho}_A)(t) \equiv \frac{1}{2}||\hat{\rho}_A(t) - \hat{\rho}_A^\mathrm{th}||_1,
\end{equation}
where $\hat{\rho}_A(t)$ is the reduced density matrix over $A$ spins at time $t$, $\hat{\rho}_A^\mathrm{th}$ is the thermal density matrix of the final equilibrium state, and $||\hat M||_1 \equiv \Tr(\sqrt{\hat M^\dagger \hat M})$ is the trace norm of an operator $\hat M$. The thermal density matrix associated with a state $\ket{\psi}$ takes the form $\hat \rho^\mathrm{th} = (1/Z)e^{-\beta \hat H}$, where $\beta$ is the inverse of the effective temperature corresponding to the state $\ket{\psi}$ and $Z=\mathrm{Tr}\, e^{-\beta\hat H}$. This effective temperature is determined self-consistently by requiring 
\begin{equation}\label{eq:effectivetemp}
    \langle \psi| \hat H |\psi\rangle = \Tr(\hat H\rho^\mathrm{th}).
\end{equation}

We expect that our results should not depend on the choice of measure used to analyze the QME. To demonstrate this, we calculated the trace distance for the states defined in Eq.~(9) of the main text with $\theta=\pi/4$ and $\theta=\pi/8$, evolved under the XYZ Hamiltonian with $J_x=-0.5$, $J_y=-1.5$, $J_z=-0.75$ and $\alpha=4$, with varying magnetic field strengths $h_z$. The states were evolved using finite-system TDVP for MPS at system size $N=100$ and the reduced density matrix is evaluated for $N_A=4$ spins in the middle of the chain. For simplicity, $\hat\rho^\mathrm{th}$ was obtained in a system of $14$ spins using exact diagonalization techniques.  Figure~\ref{fig:tracedist} shows the trace distance ratio, $\Delta^{\theta=\frac{\pi}{4}}(\hat{\rho}_A) / \Delta^{\theta=\frac{\pi}{8}}(\hat{\rho}_A)$, which is the analog of the entanglement asymmetry ratio in Fig.~2 of the main text. At weak fields, the trace distance ratio shows no QME, while at higher field strengths the QME starts to occur. 

Comparing this to Fig.~2(e) of the main text, we see that the trace distance shows the same effect of increasing field strength leading to the occurrence of the QME. This demonstrates that, for finite systems with short range interactions, the trace distance can also be used to probe how the QME responds to symmetry breaking and emergent symmetries, even though it is not explicitly related to symmetry like entanglement asymmetry. However, for finite systems with longer-range interactions, the effects of SSB are not consistently seen, with convergence to the thermodynamic limit worsening as $\alpha$ decreases. As a result, finite systems with longer-range interactions cannot be consistently compared with systems in the thermodynamic limit, as discussed further in Sec.~\ref{app::scaling}.

 \begin{figure}[tb]
    \centering
    \includegraphics[width=0.85\linewidth]{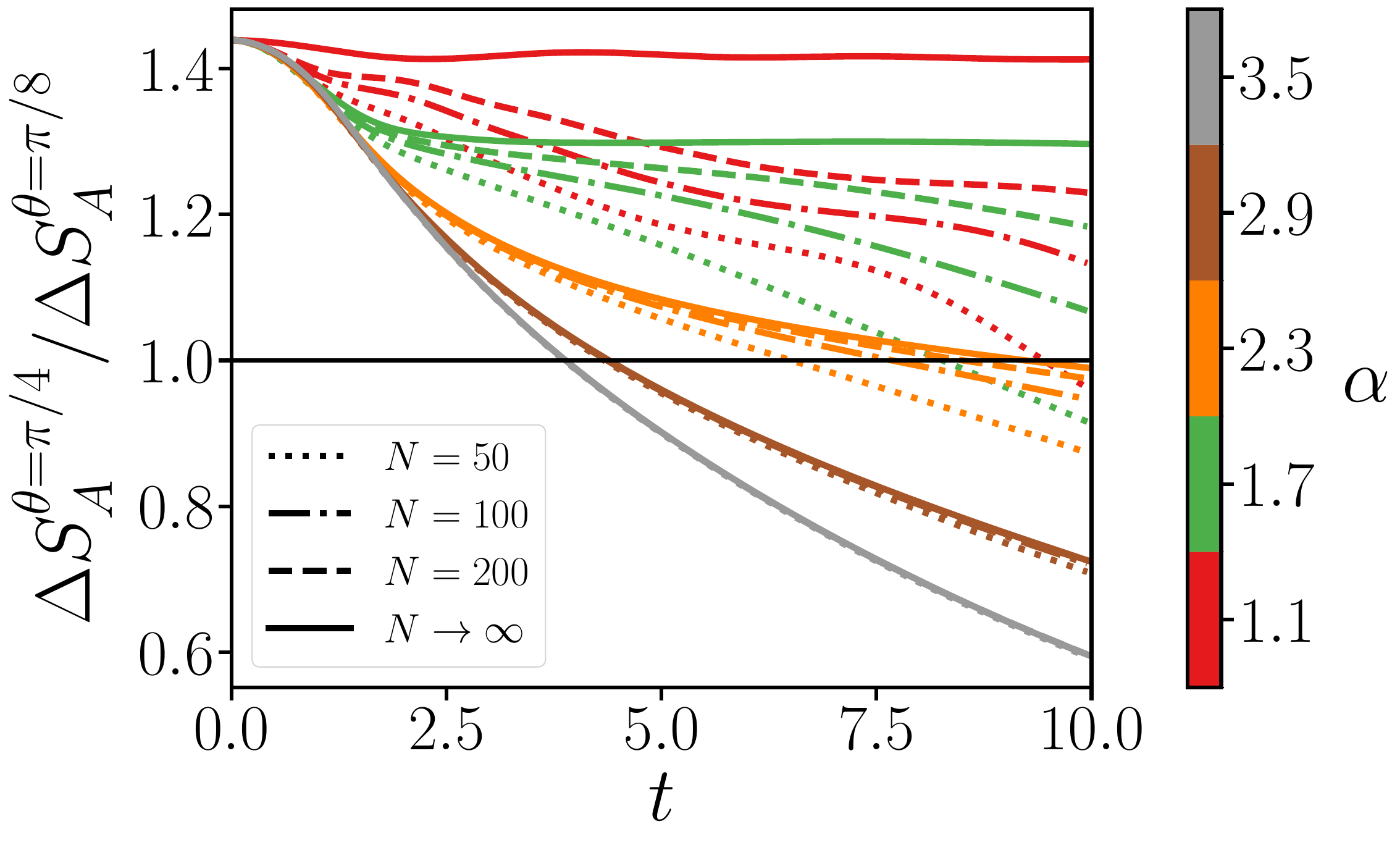}
    \caption{Ratio of the entanglement asymmetry for the states in Eq.~(9) of the main text, with $\theta_1=\pi/4$ and $\theta_2=\pi/8$, evolved under the XXZ Hamiltonian [Eq.~(3) of the main text] with $J_x=J_y=-1$ at $J_z=-0.75$ for various values of $\alpha$ (color bar). For each value of $\alpha$, we show data for increasing finite system sizes $N=\{50,100,200\}$ and the $N \rightarrow \infty$ limit. The ratio crossing $1$ indicates that the QME has occurred. Dynamics at finite $N$ were simulated using TDVP for MPS, with the finite systems results using maximum bond dimension $\chi=100$ and time step $\delta t=0.05$, while the $N\to\infty$ data are results of iMPS simulation with maximum bond dimension $\chi=128$ and time step $\delta t=0.025$.
    }
    \label{fig:systemsizescaling}
\end{figure}

\section{Long-range interactions in DMRG}\label{app::dmrg}

A consequence of power-law decaying interactions is that the required bond dimension for a full matrix product operator (MPO) representation of the Hamiltonian is too large to efficiently simulate. To avoid this issue, we approximate the power-law interactions as a sum of exponential terms~\cite{Pirvu2010}. This was done by fitting the parameters $a_k$ and $b_k$ in the expression
\begin{equation}\label{eq:sumofexp}
    f(n) = \sum_{k=1}^K a_ke^{b_kn}
\end{equation}
to the domain $\{n:n\in[0,N-2]\}$ and the range $\{\frac{1}{n}:n\in[1,N-1]\}$, where $K$ is the number of exponential terms used to approximate the power law. With this choice of domain and range, the value $f(0)=\sum_{k=1}^K a_k$ will represent the nearest-neighbor interaction term $\frac{1}{2}$. Using this approximation, the bond dimension of the XYZ Hamiltonian MPO becomes $2+3K$. For all of the results presented in the paper, the interactions were fitted to a sum of $K=8$ exponential terms.

To justify the choice of $K$, we checked the convergence of our results by repeating the calculation  with $K = 6$, $8$, and
$10$ exponential terms. Figure~\ref{fig:numExp}(a) shows the
entanglement asymmetry ratio $\Delta S_A^{\theta=\pi/4}/\Delta
S_A^{\theta=\pi/8}$ [cf. Fig.~3(a) of the main text], evolved under
the same XXZ Hamiltonian, for representative values of $\alpha$
(color bar), comparing $K=6$ (dotted), $K=8$ (dashed), and $K=10$
(solid). The $K=8$ and $K=10$ curves lie essentially on top of one
another across the full range of $\alpha$ and times considered,
while the $K=6$ curves show a small but visible deviation at late
times for intermediate $\alpha$. Figure~\ref{fig:numExp}(b)
quantifies this by plotting the relative error of the $K=8$
approximation used throughout this work,
\begin{equation}\label{eq:relerr}
    \text{Error}(t) = \frac{|\Delta S_A^{K=10}(t) - \Delta S_A^{K=8}(t)|}{\Delta S_A^{K=10}(t)},
\end{equation}
as a function of time. The error remains below $\sim 1\%$ throughout
the evolution for all $\alpha$. This confirms that $K=8$ is
sufficiently converged and does not introduce systematic errors
affecting our conclusions.
    
Due to finite-size effects, certain regions of the phase diagram showed significant oscillations in entanglement entropy depending on the location of the bond defining the bipartition. As a result, if the entanglement entropies used to calculate $c_\mathrm{eff}$ were taken from an even bond in one system and an odd bond in the other, the difference would compare a peak of the oscillation in one system with a trough of the oscillation in the other, resulting in an erroneously large estimate of $c_\mathrm{eff}$. To solve this problem, when comparing two system sizes with one having a bipartition on an even bond and the other on an odd bond, the entanglement entropy would instead be taken for one of the systems by moving the bond to $N/2 + 1$. This ensures that the comparison is made between two peaks (or two troughs, depending on which of the two systems is chosen for this shift). This correction works reliably for sufficiently large systems, where the entanglement entropy will be slowly varying about the peak, such that the shift gives a good approximation.

The DMRG was performed using the ITensor library \cite{ITensor, fishman2025itensor-r0.3}. The DMRG was performed with a max bond dimension of $\chi=200$ and cutoff of $10^{-14}$ with $10$ DMRG sweeps. For the Krylov eigensolver the max Krylov dimension was set to $20$, the max number of eigensolver iterations was set to $300$ and the eigensolver tolerance was set to $10^{-14}$. 

\begin{figure}[tb]
    \centering
    \includegraphics[width=0.85\linewidth]{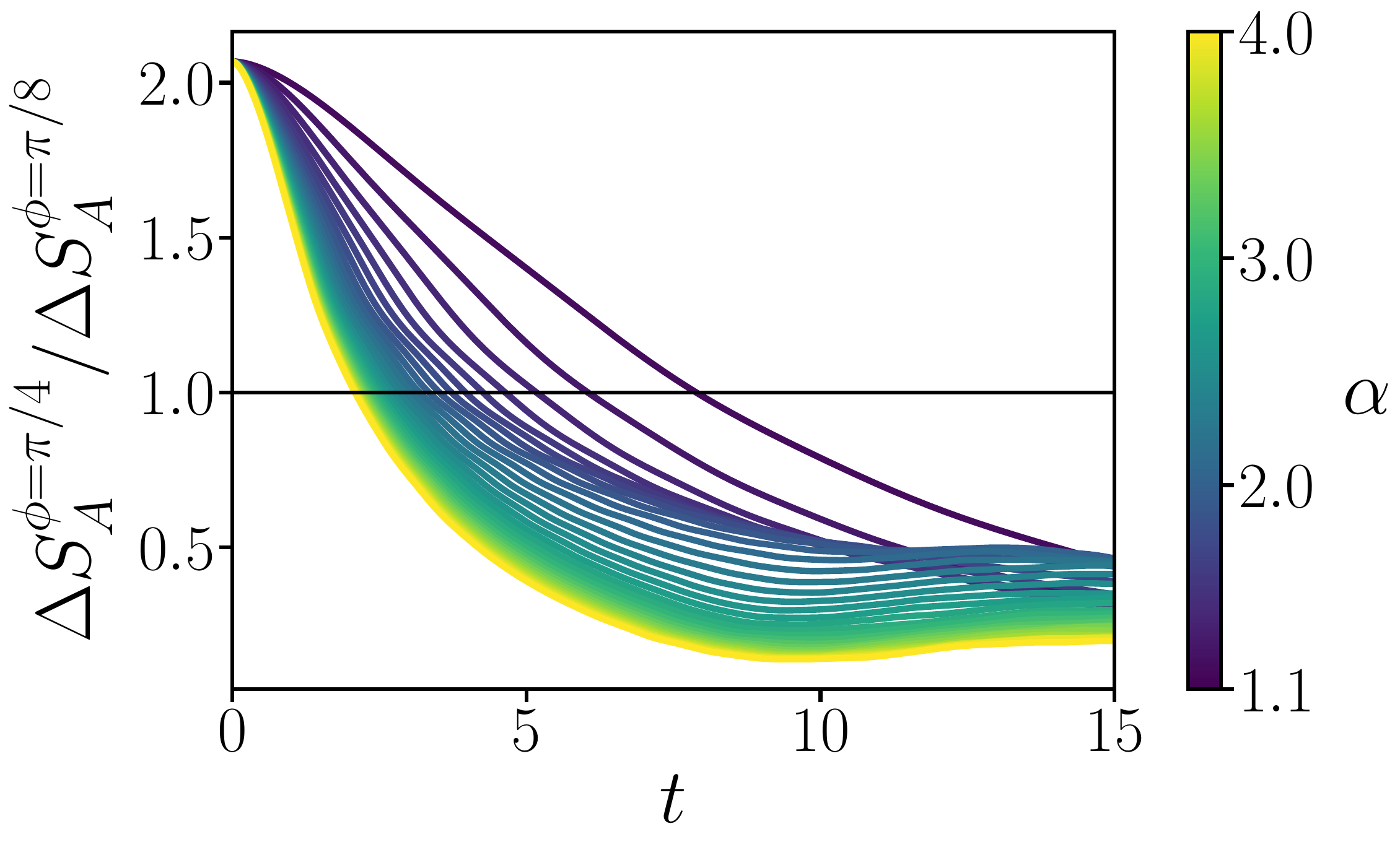}
    \caption{Ratio of the SU(2) entanglement asymmetry for the states in Eq.~(11) of the main text, with $\phi_1=\pi/4$ and $\phi_2=\pi/8$, evolved under the XXZ Hamiltonian [Eq.~(3) of the main text] with $J_x=J_y=J_z=-1$ for various values of $\alpha$ (color bar). The ratio crossing $1$ indicates that the QME has occurred. All data were obtained for a subsystem of $N_A=4$ spins in an infinite chain using TDVP with translation-invariant iMPS of bond dimension $\chi=128$ and a timestep $\delta t=0.025$.}
    \label{fig:su2plot}
\end{figure}

\section{Finite-size scaling}\label{app::scaling}

In this section, we present a more systematic analysis of finite-size scaling for the QME in the XXZ model. As the range of interactions increases, the convergence toward the infinite model becomes slower. This poses a challenge when analyzing the halting of the QME, since it depends on SSB---a phenomenon that occurs in the thermodynamic limit. In the simulations presented in the main text, this issue was circumvented by using TDVP for iMPS, which works directly in the thermodynamic limit. Nevertheless, quantum simulations or small-scale experiments are necessarily limited to finite numbers of spins, which could be far from the thermodynamic limit. 

To explore the impact of finite size, we perform TDVP for finite systems with time step $\delta t = 0.05$, max bond dimension $\chi=100$, cutoff $10^{-12}$ and max Krylov dimension $10$. In Fig.~\ref{fig:systemsizescaling}, we show the entanglement asymmetry ratio for states in Eq.~(9) of the main text, with $\theta_1=\pi/4$ and $\theta_2=\pi/8$, evolved under the XXZ Hamiltonian [Eq.~(3) of the main text] with $J_x=J_y=-1$ and $J_z=-0.75$. We show data for increasing finite system sizes and compare them to the infinite limit for various values of $\alpha$. It can be seen that even at the largest system size shown, the halting behavior of the QME seen in the infinite limit is not captured when $\alpha \lesssim 2$. The numerical analysis in finite systems is furthermore complicated by the existence of $O(N^{1/2})$ system-size dependent oscillations due to the Anderson tower of states associated with SSB~\cite{Anderson1952,Tasaki2019}, which requires both large systems and large evolution times to ascertain whether the asymmetry ratio eventually crosses unity. For these reasons, in the main text we chose to work directly in the thermodynamic limit, allowing to more reliably identify signatures of SSB. 

\begin{figure}[tb]
\centering
\includegraphics[width=0.85\linewidth]{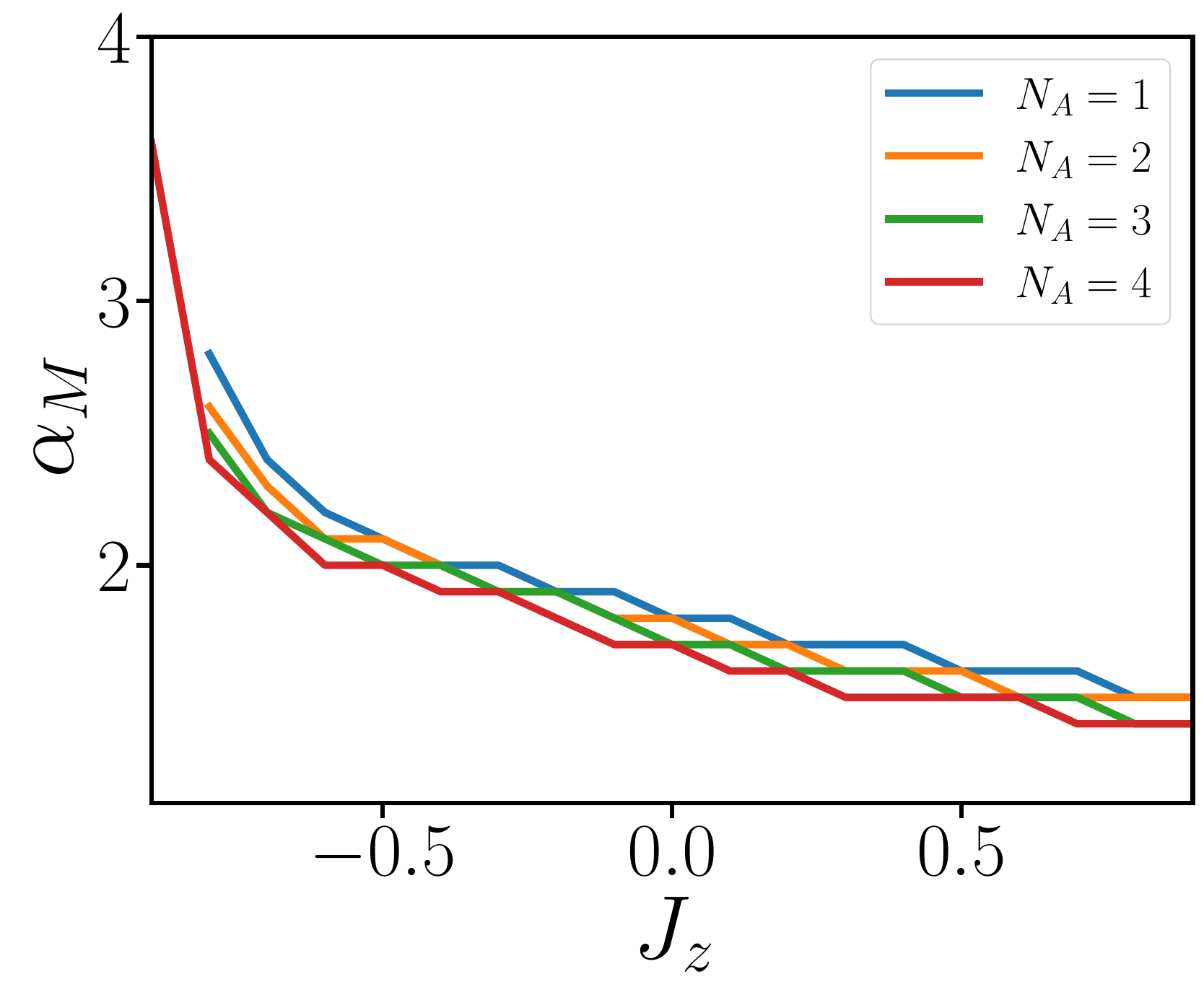}
\caption{The dependence of $\alpha_M$ on $J_z$ for several subsystem sizes $N_A$ indicated in the legend. Data is for the tilted ferromagnetic states with $\theta=\pi/4$ and $\theta=\pi/8$ in the XXZ model. For $N_A=1-3$, we omit the data point for $J_z=-0.9$ as the Mpemba time did not diverge for the range of values $\alpha\in[1.1,4]$ we checked. }\label{fig:varyingsubsystem}  
\end{figure}

Finally, even if we work in the thermodynamic limit, one may wonder how our results are affected by the size of the \emph{subsystem}. We repeat the analysis from Fig.~3(c) of the main text using a range of subsystem sizes  $N_A\in\{1,4\}$. Fig.~\ref{fig:varyingsubsystem} shows the value of $\alpha$ at which the Mpemba time $\tau_M$ diverges for each subsystem size. We see that, as the size of the subsystem increases, the critical $\alpha_M$ slightly shifts downwards, with somewhat longer-ranged interactions required to halt the QME. On the other hand, there is also a larger uncertainty in extracting $\alpha_M$ for smaller $N_A$ due to larger fluctuations of the entanglement asymmetry during the evolution. Ultimately, keeping in mind experimental realizations, we decided to focus on subsystems with $N_A=4$, with Fig.~\ref{fig:varyingsubsystem} showing that the results should be robust for small finite subsystems.

\begin{figure*}
    \centering
    \includegraphics[width=0.99\linewidth]{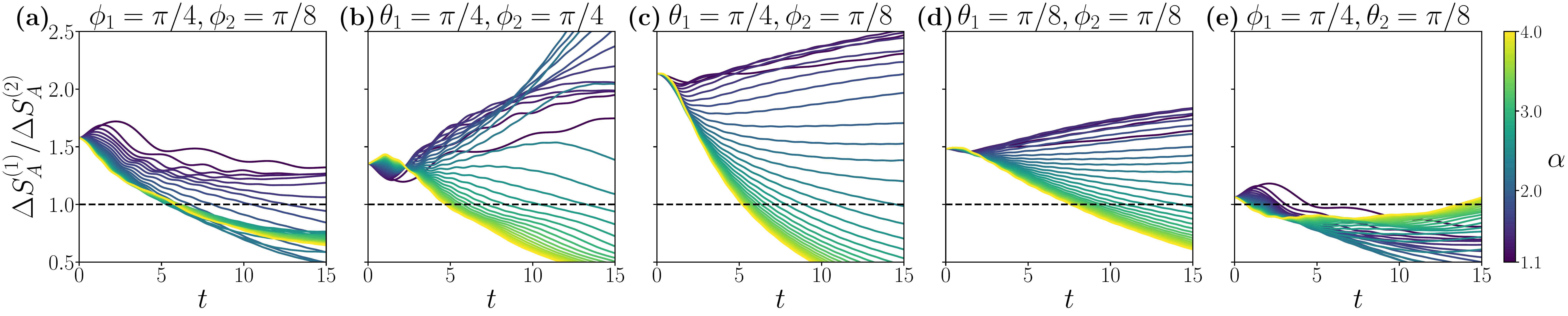}
    \caption{
    Ratio of the entanglement asymmetry for different combinations of initial states.
    (a) The initial states in Eq.~(11) of the main text, with $\phi_1= \pi/4$ and $\phi_2= \pi/8$. 
    (b)-(e): One of the initial states is given by Eq.~(9) of the main text (denoted by $\theta$) and the other by Eq.~(11) of the main text (denoted by $\phi$). In all the panels, the states are evolved under the XXZ Hamiltonian [Eq.~(3) of the main text] with $J_x=J_y=-1$ and $J_z=-0.75$ for various values of $\alpha$ (color bar). The ratio crossing $1$ indicates that the QME has occurred. The results were obtained for a subsystem of $N_A=4$ spins using TDVP with translation-invariant iMPS of bond dimension $\chi=128$ and a timestep $\delta t=0.025$.
    }
    \label{fig:2site phase diagram}
\end{figure*}

\section{Mpemba effect at the SU(2) point}\label{app::SU2 results}

In this section, we focus on the effective Heisenberg description of the XYZ model, which requires $h_z \gg |J_x-J_y|$ similar to the effective XXZ model, and has additional requirement of $h_z \gg | (J_x+J_y)/2-J_z|$. This stronger condition extends the emergent U(1) symmetry to an emergent SU(2) symmetry, leading to effective isotropic spin-spin couplings in the $x$, $y$, and $z$ directions. Following the approach of the main text,  we study this regime through analyzing the SU(2) symmetric Heisenberg model to understand how the QME depends on the range of interaction. In contrast to most of the main text, here we will be discussing the QME with respect to the SU(2) symmetry, where we use the same definition of the QME and entanglement asymmetry $\Delta S_A$ from the main text, where we now assume that the generator of the global SU(2) symmetry $\hat Q$ is such that
\begin{equation}\label{eq:su2generator}
    \hat Q_A = \sum_{\nu\in\{x,y,z\}}(\sum_{i\in A} \hat \sigma^\nu_i)^2.
\end{equation}
Similar to the U(1) case, we define the SU(2) entanglement asymmetry using the projectors $\Pi_q$ onto the eigenspaces of $\hat Q_A$ with charge $q\in\mathbb{Z}$, which give the projected reduced density matrix $\rho_{A,Q}$ introduced in the main text. An important detail is that the product initial states in Eq.~(9) of the main text are eigenstates of the SU(2) generator and thus SU(2)-symmetric with $\Delta S_A(\theta, t)=0$, hence they will remain invariant under evolution with an SU(2) symmetric Hamiltonian. Instead, we use the tilted N\'eel states in Eq.~(11) of the main text, which are parameterized by $\phi$ and break SU(2) symmetry. 

Fig.~\ref{fig:su2plot} shows the entanglement asymmetry ratio $\Delta S_A^{(1)}/\Delta S_A^{(2)}$ for these states in Eq.~(11) of the main text, with $\phi_1=\pi/4$ and $\phi_2=\pi/8$, evolved under the XXZ Hamiltonian [Eq.~(3) of the main text] with $J_x=J_y=J_z=-1$, i.e. the long-range Heisenberg model, for various values of $\alpha$ (color bar). We find that the QME persists at all interaction ranges, in contrast to the U(1) QME which is halted by sufficiently long-ranged interactions. This is due to the lack of SU(2) SSB in the Heisenberg model: even as the interaction range increases, symmetry is not broken. Without SSB there is still symmetry restoration even with long-range interactions. This highlights the importance of symmetry breaking in halting the QME. Without symmetry breaking, the dynamical phase diagram in Fig.~4 of the main text does not apply to the SU(2) QME in this system. To realize the right column of the dynamical phase diagram for the SU(2) QME a system with SU(2) SSB is required, such as the long-range spin $1/2$ Heisenberg model in $d=2$.

\section{Other initial states}\label{app::other state phase diagram}

In this section, we present further numerical evidence for the QME phase diagram in Fig.~4 of the main text by considering different combinations of initial states introduced in the main text. There, we introduced two families of experimentally-preparable initial states, Eq.~(9) and Eq.~(11), but so far we have primarily considered the QME between states belonging to the first family. The energy density of these states with respect to the XXZ Hamiltonian [Eq.~(3) of the main text] were calculated using the formula:
\begin{equation}
    \frac{E_{\theta}(\alpha)}{N} = \frac{1}{2}\Big(J_x \sin^2(2\theta) + J_z \cos^2(2\theta)\Big)\,,
\end{equation}
for the tilted product states in Eq.~(9), and
\begin{eqnarray}
\notag \frac{E_{\phi}(\alpha)}{N} &=& \frac{1}{2^{2+\alpha}}\Big[J_x\sin^2(2\phi) \\ &+& J_z\Big(1+\cos(2\phi)(\cos(2\phi)-2 + 2^{1+\alpha})\Big)\Big],\quad
\end{eqnarray}
for the tilted N\'eel states in Eq.~(11). These formulae were calculated for the thermodynamic limit and $E_{\min}/N$ and $E_{\max}/N$ were estimated  to be $-0.503$ and $0.466$ respectively. This was calculated using DMRG for a system of $N=200$ spins with $J_x=J_y=-1$, $J_z=-0.75$ and $\alpha=2$. Figure~\ref{fig:2site phase diagram}(a) demonstrates that similar results are obtained when considering the second family. Figure~\ref{fig:2site phase diagram}(a) shows the ratio of the entanglement asymmetry, for the states with $\phi_1=\pi/4$ and $\phi_2=\pi/8$, evolved under the XXZ Hamiltonian with $J_x=J_y=-1$ and $J_z=-0.75$ at various values of $\alpha$. Similar to the tilted product initial states [Fig.~3(a) of the main text], we observe that the QME occurs at short interaction ranges and is halted as the range of interaction increases. From this, we infer that the QME diverges at a smaller value of $\alpha$ compared to the tilted product initial states [cf. Fig.~3(b) in the main text], due to their higher energy density which requires a longer range of interactions to stabilize the U(1) SSB. 

Figures~\ref{fig:2site phase diagram}(b)-(e) show the ratio of the entanglement asymmetry for different pairs of states, one state from the family of tilted product states in Eq.~(9) of the main text (indicated by $\theta$) and one state belonging to the tilted N\'eel family in Eq.~(11) of the main text (indicated by $\phi$). Overall, the behavior is consistent with the dynamical phase diagram in Fig.~4 of the main text. Additionally, for these combinations of states, the Mpemba time diverges at a similar value of $\alpha$ compared to the combination of the two tilted product initial states. However, it should be noted that for Figures~\ref{fig:2site phase diagram}(b) and (c) a long enough timescale was not reached to accurately determine the $\alpha$ at which the Mpemba time diverges, as can be seen with the entanglement asymmetry ratios near and above $1$ at the end of the timescale considered still tending downwards.

Another feature of Figures.~\ref{fig:2site phase diagram}(b)-(d) is that the state with a lower energy-density appears at the top of the entanglement asymmetry ratio (to ensure the initial ratio is greater than unity). The lower energy-density state will undergo SSB at a shorter range of interactions compared to higher energy-density states. Consequently, there exists an intermediate range of $\alpha$ where the state at the top of the ratio undergoes SSB whilst the state at the bottom does not, meaning that the state that started at a higher initial entanglement asymmetry will plateau whilst the entanglement asymmetry of the other state will return to $0$. For these $\alpha$ values we observe an initial dip in the ratio followed by an increase caused by the simultaneous plateau of the state with higher entanglement asymmetry and decrease in the entanglement asymmetry of the state that is lower. This is also due to the different timescales on which these states evolve, with the state at a higher entanglement asymmetry quickly plateauing, and the other state still decreasing in entanglement asymmetry at comparably late times. This effect is absent when comparing states of similar energy density, where both entanglement asymmetries plateau or vanish on similar timescales at similar interaction ranges.

On the other hand, Fig.~\ref{fig:2site phase diagram}(e) does not follow the same trend, which is attributed to the initial entanglement asymmetry ratio being too close to unity. This suggests that the phase diagram in Fig.~4 of the main text is expected to hold for states whose initial entanglement asymmetry ratio is sufficiently larger than unity. For short interaction ranges, the ratio in Fig.~\ref{fig:2site phase diagram}(e) crosses unity rapidly, but at late times it crosses again. This unusual late-time behavior is pushed back as $\chi$ increases showing that it is an artifact of the limited bond dimension used, a common issue affecting TDVP at longer times.

\begin{figure*}[tb]
    \centering
    \includegraphics[width=0.99\linewidth]{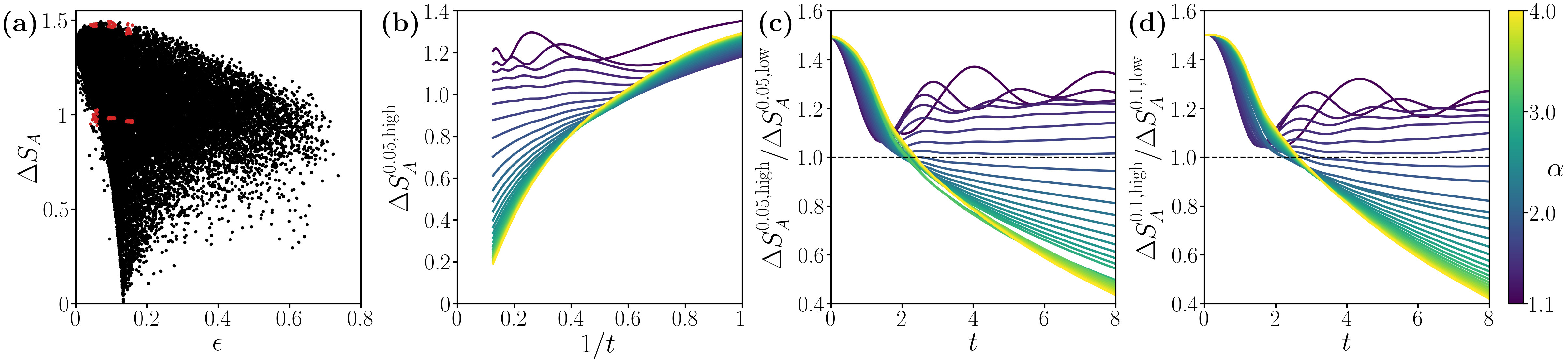}
    \caption{
    Halting of the QME for random MPS initial states.
    (a): The entanglement asymmetry of Haar random, $\chi=2$ MPS states at different normalized energy density $\epsilon$. Highlighted in red are twenty initial states chosen close to $\epsilon=0.05,0.1,0.15,0.2$ at high and low entanglement asymmetry.  
    (b): The average entanglement asymmetry of the twenty high entanglement $\epsilon=0.05$ states as a function of inverse time $t$ for $\chi=180$.
    (c)-(d): The entanglement asymmetry ratio for the $\epsilon=0.05$ and $\epsilon=0.1$ states, demonstrating the halting of the QME.
    }
    \label{fig:randstates}
\end{figure*}

\section{Halting of the QME for random states}\label{app::random_states}

Since the QME emerges through a comparison of $\Delta S_A$ for two states, it is important to verify whether this behavior persists for broader classes of initial states. The long-range XYZ spin chain studied in this work is non-integrable for any finite $\alpha$, hence it is expected to satisfy the ETH. From this perspective, the only property of the initial state that should determine its late-time dynamics is its energy density, while short-time dynamics may still depend on additional microscopic details.

To test the sensitivity of the QME, we consider an ensemble of Haar-random MPS states. Such states can be generated from a Haar-random unitary $U \in U(d\chi)$, where $d=2$ is the physical dimension and $\chi=2$ is the bond dimension. Selecting the first $\chi$ columns of $U$ yields a $d\chi \times \chi$ matrix, which can be reshaped into a $\chi \times d \times \chi$ tensor defining the MPS. Since we work directly in the thermodynamic limit, these states are described by local random MPS tensors, which are then infinitely repeated, resulting in an overall translation-invariant state. The corresponding values of $\Delta S_A$ and the normalized energy density $\epsilon$ [Eq.~(10) of the main text] for an ensemble of $25{,}000$ Haar-random MPS states with $\chi=2$ are shown in Fig.~\ref{fig:randstates}(a). From this ensemble, we select $20$ states with relatively high and low values of $\Delta S_A$ at fixed energy densities $\epsilon = 0.05, 0.1,$ and $0.15$, shown in red in Fig.~\ref{fig:randstates}(a).

In Fig.~\ref{fig:randstates}(b) we check to what extent the selected Haar-random initial states reproduce the behavior observed in the main text. We plot $\Delta S_A$ as a function of $1/t$, averaged over the high-$\Delta S_A$ states at $\epsilon=0.05$. As $\alpha \rightarrow 1$, $\Delta S_A$ saturates to a finite value, signaling the emergence of SSB. As for the initial states discussed in the main text, this SSB results in a halting of the QME at sufficiently small $\alpha$.

Analogous behavior is further illustrated in Fig.~\ref{fig:randstates}(c) for $\epsilon=0.05$ and Fig.~\ref{fig:randstates}(d) for $\epsilon=0.1$, which demonstrate that the halting of the QME persists for generic initial states across different energy densities. Taken together, these results strongly suggest that the phenomena reported in the main text are generic for a broad class of low-entangled initial states, and do not rely on special properties of the (anti)ferromagnetic product states.

\begin{figure}[bt]
    \centering
    \includegraphics[width=\linewidth]{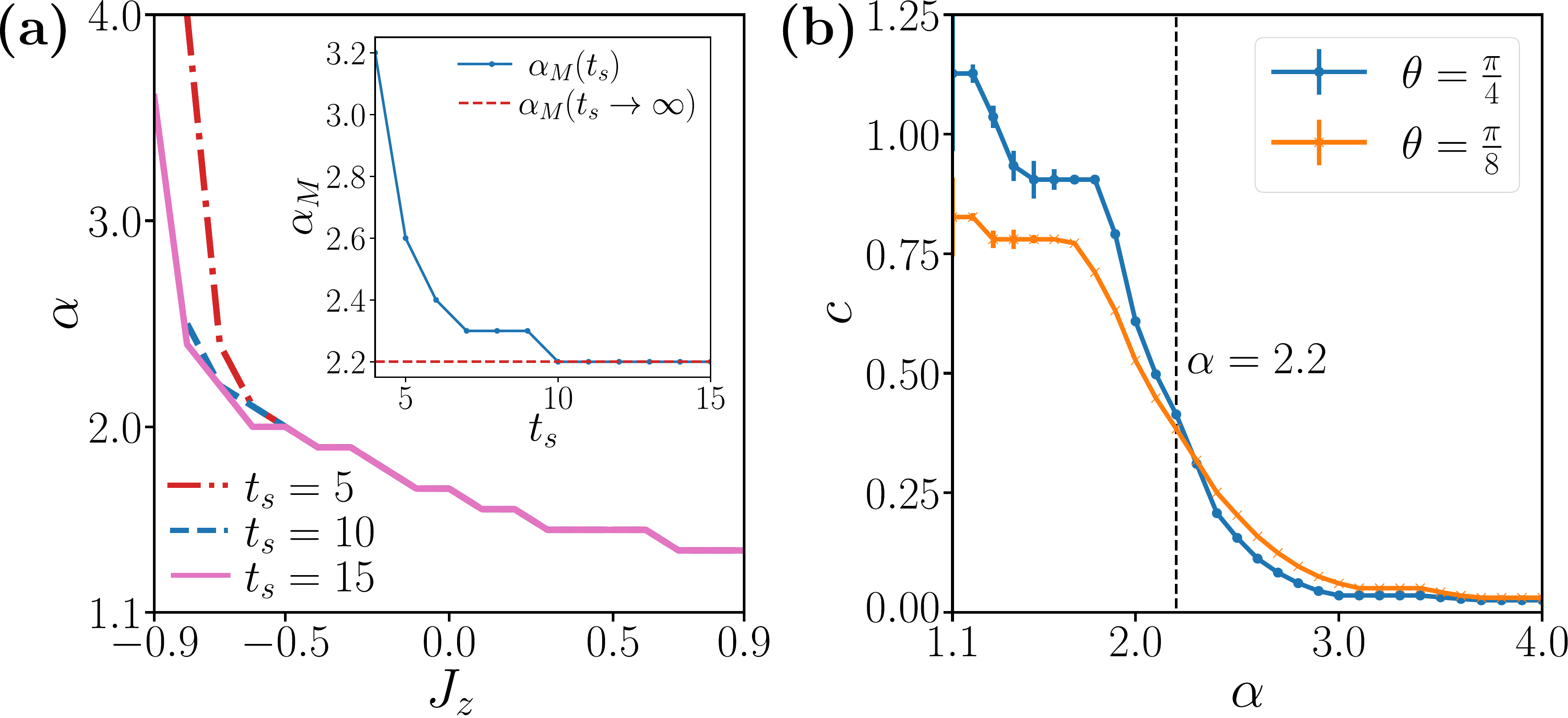}
    \caption{(a) The interaction range $\alpha_M$ at which the Mpemba time diverges, plotted for different maximum simulation times $t_s$. Inset shows the convergence of $\alpha_M$ at $J_z=-0.75$ with increasing the maximum simulation time. The dashed line is the value obtained from infinite-time extrapolation using the function $\Delta S_A^\theta = a \exp(-bt) + c$.
    (b) The residual value $c$ of the fitted entanglement asymmetry at $t\to\infty$ as a function of $\alpha$. Data is for the titled ferromagnetic states with  $\theta=\pi/4$ and $\theta=\pi/8$, at fixed $J_z=-0.75$, as a function of $\alpha$. The crossing point (dashed line) marks the threshold for the onset of the QME.}
    \label{fig:inftimeextrapolation}
\end{figure}

\section{Infinite-time extrapolation}\label{app::TDVP_conv}

The analysis presented in the main text is dependent on the maximum simulation time. This is due to the use of iMPS which was necessary to directly observe SSB by simulating the dynamics in the thermodynamic limit $N\rightarrow\infty$. To confirm that our conclusions hold as the simulation time $t_s\rightarrow\infty$, here we perform a late-time extrapolation of the iMPS results in Fig.~3(a) of the main text, and investigate whether in the late-time limit the QME halts with increasing range of interaction.

Figure~\ref{fig:inftimeextrapolation}(a) shows how the line $\alpha_M$ moves as we increase the maximum simulation time $t_s$. For all $J_z>-0.5$, the value of $\alpha_M$ quickly converges with $t_s$. We only observe slower convergence in the vicinity of $J_z=-1$, which is due to the nearby SU(2)-symmetric point, where the chosen initial states become eigenstates of the Hamiltonian and the Mpemba time diverges. As a result, longer simulation times are required in this region to capture when the QME halts. Nevertheless, even very close to the SU(2) point, the data for $t_s=10$ and $t_s=15$ are almost on top of each other, illustrating the robustness of our $\alpha_M$ estimates.

Furthermore, we fitted the decay of the entanglement asymmetry at $J_z=-0.75$ using $\Delta S_A^\theta = a \exp(-bt) + c$, where $a$,$b$ and $c$ are fitting parameters~\footnote{We note that we have checked a power-law fit, which gave much larger residual errors compared to the exponential decay}. The inset of Fig.~\ref{fig:inftimeextrapolation}(a) shows how $\alpha_M$ converges to the infinite-time extrapolation value as we increase the simulation time $t_s$. This demonstrates that the dynamics at $J_z=-0.75$ have converged to the infinite-time value. Similar results are obtained for other values of $J_z\gtrsim -0.9$ (data not shown). Moreover, in Fig.~\ref{fig:inftimeextrapolation}(b) we see that, at large $\alpha$, the value of $c$ for $\theta=\pi/4$ is less than that for $\theta=\pi/8$, which implies that there is a QME in the limit $t\rightarrow\infty$. The threshold value for the QME is the crossing point at $\alpha=2.2$, which matches the $\alpha_M$ value shown in Fig.~3(c) of the main text. We note that the same fitting analysis was also done on the results in Fig.~2(f), which supported our conclusions that in the late time limit the QME does not occur.

\bibliography{references}